\documentclass[aps,pre,reprint,onecolumn,notitlepage,superscriptaddress]{revtex4-2}
\usepackage{amssymb,amsmath,color,mathtools,subeqnarray,bm,mathtools,upgreek}
\usepackage[colorlinks=true,linkcolor=blue]{hyperref}
\usepackage[sfdefault=cmbr]{isomath}
\usepackage{graphicx}
\usepackage{epstopdf, epsfig}
\usepackage{xcolor}
\usepackage{subfig}
\usepackage{comment}

\def \t{\tensorsym}

\begin{document}

\title{Active spheroids in viscosity gradients}

\author{Jiahao Gong}%
\affiliation{%
 Department of Mathematics,\\
University of British Columbia, Vancouver, BC, V6T 1Z2, Canada
}%

\author{Vaseem A. Shaik}
\affiliation{
Department of Mechanical Engineering,\\
University of British Columbia, Vancouver, BC, V6T 1Z4, Canada
}%

\author{Gwynn J. Elfring}
 \email{gelfring@mech.ubc.ca}
\affiliation{%
 Department of Mathematics,\\
University of British Columbia, Vancouver, BC, V6T 1Z2, Canada
}%
\affiliation{
Department of Mechanical Engineering,\\
University of British Columbia, Vancouver, BC, V6T 1Z4, Canada
}%

\begin{abstract}
In this paper, we explore the hydrodynamics of spheroidal active particles in viscosity gradients. This work provides a more accurate modeling approach, in comparison to spherical particles, for anisotropic organisms like \textit{Paramecium} swimming through inhomogeneous environments, but more fundamentally examines the influence of particle shape on viscotaxis. We find that spheroidal squirmers generally exhibit dynamics consistent with their spherical analogs, irrespective of the classification of swimmers as pushers, pullers, or neutral swimmers. However, the slenderness of the spheroids tends to reduce the impact of viscosity gradients on their dynamics; when swimmers become more slender, the viscosity difference across their body is reduced, which leads to slower reorientation. We also derive the mobility tensor for passive spheroids in viscosity gradients generalizing previous results for spheres and slender bodies. This work enhances our understanding of how shape factors into the dynamics of passive and active particles in viscosity gradients, and offers new perspectives that could aid the control of both natural and synthetic swimmers in complex fluid environments. 
\end{abstract}

\maketitle

\section{Introduction}
Active particles, which include both biological organisms and synthetic particles, have the capability to convert stored energy to directed movement \citep{Schweitzer2007}. A large number of active particles can form a dynamic system commonly referred to as active matter. The active constituents in active matter can span a wide range of scales from nanorobots and microswimmers to larger organisms like birds, fish, and even humans \citep{Viscek2012, Marchetti2013}. In this study, we focus on micron-sized active particles. The widespread existence of microorganisms in natural settings, combined with substantial advancements in microfluidic experimental techniques, has led to an explosion of research focusing on the motion of small active particles, both biological and synthetic, in viscous fluids \citep{Lauga2009, Elgeti2015, Bechinger2016}.  

Active particles often exist within gradients of a variety of physical quantities such as heat, light \citep{Jekely2009}, or chemicals \citep{Moran2017}, and often respond by reorienting themselves to swim up or down these gradients, a behavior known as taxis. For instance, \textit{E. coli} is found to display chemotaxis in gradients of oxygen, galactose, glucose, aspartic acid, threonine, or serine \citep{Adler1966}. Meanwhile, the photophobic behavior of \textit{E. coli} can be used to `paint' with bacteria by selective exposure to light \citep{Arlt2018}.  Here we focus on taxis due to environments that are \textit{mechanically} inhomogeneous, specifically where the viscosity is spatially varying. Viscosity gradients can be found in nature when properties of the fluid such as temperature, salinity, or even suspended substances are spatially varying. As an example, numerous coral species secrete mucus that builds up on the sea's surface, leading to areas with differing viscosities where marine microorganisms navigate \citep{Wild2004, Guadayol2021}. It has also been shown that the movement and distribution of intestinal bacteria is influenced by viscosity variations in the mucus layer \citep{Swidsinski2007}.

Previous experimental studies have observed that several microorganisms demonstrate apparent viscotaxis. For example, \textit{Leptospira} and \textit{Spiroplasma} are observed to propel up viscosity gradients \citep{Kaiser1975, Petrino1978, Daniels1980, Takabe2017}. In contrast, \textit{E. coli} have been observed to swim down the viscosity gradients \citep{sherman1982}. \textit{Chlamydomonas reinhardtii}, a type of green microalgae, demonstrates complex behavior in viscosity gradients: it accumulates in high-viscosity regions when gradients are weak, but reorients towards low-viscosity regions in strong gradients \citep{Stehnach2021}. When interacting with sharp viscosity gradients, this same alga displays dynamics analogous to the refraction of light, as observed experimentally \citep{Coppola2021} and modeled theoretically \citep{Gong2023}.

Recently, it was demonstrated that a purely hydrodynamic mechanism can lead to viscotaxis \citep{Liebchen2018}. In that work, active particles were modeled as interconnected spheres propelled by a fixed thrust in weak viscosity gradients. These particles were shown to display positive viscotaxis due to an imbalance in viscous drag acting on different spheres. Later work included the effect of viscosity variations on thrust using the spherical squirmer model where the particle activity responsible for generating thrust is represented as a surface slip velocity \citep{lighthill1952,blake1971}. It was shown that hydrodynamic interactions between the active slip conditions on the squirmer's surface and the fluid with spatially varying viscosity generally leads to negative viscotaxis \citep{Datt2019, Shaik2021, Gong2023}. The dynamics of a spherical squirmer in spatially varying viscosity that results from nonuniform distribution of nutrients  has also been explored \citep{Shoele2018}. And recently, the scallop theorem \citep{Purcell1977} was shown to hold in viscosity gradients \citep{EsparzaLopez2023}.

While previous work has focused on spherical squirmers, the influence of particle shape on viscotaxis has yet to be investigated. Previous studies using a two-dimensional swimming sheet have shown speed increases when it moves either along or against gradients \citep{Dandekar2020}. More recently, it was demonstrated that viscosity gradients can introduce new forces on slender bodies, offering potential ways to control their orientation and drift \citep{Kamal2023}. Sedimenting spheroids were also shown to reorient in viscosity gradients unlike in homogeneous fluids \citep{Anand2023}. 

In order to understand the impact of particle shape on swimming in viscosity gradients, in this paper we use a prolate spheroid squirmer as a model microswimmer. Spheroidal squirmers can be used to represent ciliates with non-spherical bodies (like \textit{Tetrahymena thermophila} and \textit{Paramecium}). The model was first proposed by \citet{Keller1977} who showed that the streamlines predicted by their model closely aligned with experimental streak photographs of freely swimming and inertly sedimenting \textit{Paramecium caudatum}. Later, other researchers modified the model by adding a force-dipole mode to represent various types of swimmers, such as pushers or pullers to examine the behaviour of a single or pair of spheroidal squirmers moving in a narrow slit \citep{Theers2016}. More recent work explored the dynamics, power dissipation, and swimming efficiency, of a spheroidal squirmer in shear-thinning fluids \citep{Vangogh2022} using the reciprocal theorem, an approach similar to that which we employ in this work.

We organize this paper as follows. In \S \ref{section:Formulation}, we provide the essential mathematical details of an active prolate spheroid swimming in constant viscosity gradients. We then use the reciprocal theorem and asymptotic analysis to derive expressions for the translational and rotational velocity of the particles in \S \ref{section:reciprocal_thm}. In \S \ref{section:Passive}, we give an analytical expression for the mobility tensor of passive particles subject to an external force and/or torque.  In \S \ref{section:Active} we calculate the swimming dynamics of active prolate spheroids and compare our results with those of a spherical squirmer. In \S \ref{section:Disturbance} we discuss the effect of disturbance viscosity and \S \ref{section:Conclusion} concludes the paper.

\section{Prolate spheroids in viscosity gradients \label{section:Formulation}}
\begin{figure}
\centering 
\includegraphics[scale = 0.45]{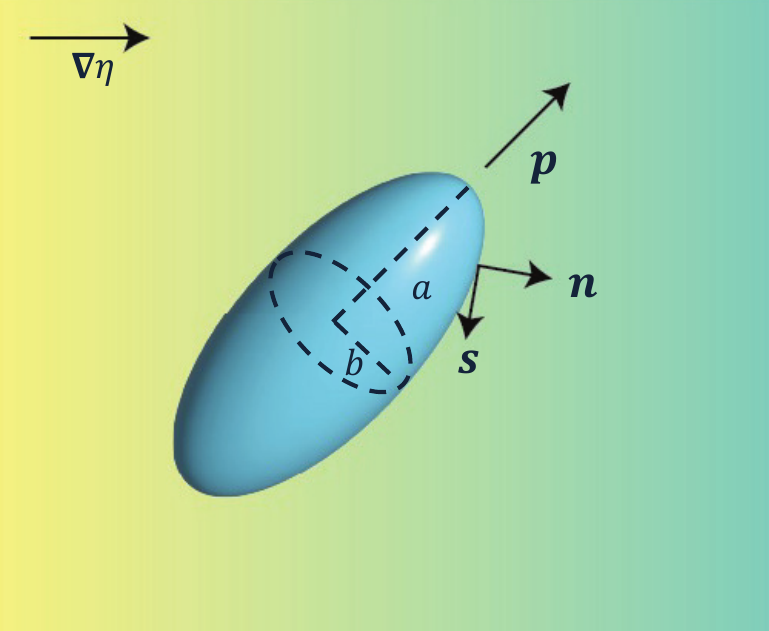}
\caption{\label{fig:schematic}Sketch of a prolate spheroidal active particle swimming in a constant viscosity gradient. $a$ and $b$ are the lengths of semi-major and semi-minor axes. The background color variations depict the viscosity variations.} 
\end{figure}

We consider a prolate spheroid particle in an otherwise quiescent Newtonian fluid. A prolate spheroid has two equatorial semi-axes of equal length and one polar longer semi-axis (see Figure~\ref{fig:schematic} for a schematic). We label the semi-major axis length $a$ and the semi-minor axis length $b$, ($b \leq a$). The eccentricity $e = \sqrt{1 - (b/a)^2}$ is a measure of the slenderness of the particle, $e=0$ being spherical, while $e=1$ is infinitely slender. The orientation of the prolate spheroid is defined as the direction $\boldsymbol{p}$ along its major axis. 

The viscosity of the fluid $\eta (\boldsymbol{x})$ is taken to be nonuniform due to spatial differences in some physical property of the fluid, such as temperature or salinity. Here we assume a constant viscosity gradient
\begin{align}
\boldsymbol{\nabla} \eta = \frac{\eta_{\infty}}{L} \boldsymbol{d},
\end{align}
where $\eta_\infty/L$ is the magnitude and $\boldsymbol{d}$ is the direction of the viscosity gradient. The size of the particle is assumed to be small compared with the macroscopic length scale of the variation of viscosity, $L$, and so we introduce a small parameter $\varepsilon = a/L \ll 1$. The viscosity gradient can then be written as $\boldsymbol{\nabla} \eta = \varepsilon \frac{\eta_{\infty}}{a} \boldsymbol{d}$.

The fluid surrounding the particle is assumed to be incompressible and Newtonian. In the limit of zero Reynolds number, the governing equations for the flow induced by particle are
\begin{eqnarray}
\boldsymbol{\nabla} \cdot \boldsymbol{u} & = & 0, \label{eqn:Stokes velocity}\\
\boldsymbol{\nabla} \cdot \boldsymbol{\sigma} & = & \boldsymbol{0}, \label{eqn:Stokes stress}
\end{eqnarray}
where $\boldsymbol{u}$ is the velocity field and $\boldsymbol{\sigma}$ is the stress tensor. The stress tensor $\boldsymbol{\sigma}$ can be written in the form
\begin{align}
    \boldsymbol{\sigma} & =  - p \boldsymbol{I} + \eta_{\infty} \dot{\boldsymbol{\gamma}} + \boldsymbol{\tau}_{NN}, \\
    \boldsymbol{\tau}_{NN} & = (\eta  (\boldsymbol{x}) - \eta_{\infty}) \dot{\boldsymbol{\gamma}},
\end{align}
where $p$ is the pressure, and $\dot{\boldsymbol{\gamma}} = {\boldsymbol{\nabla}} \boldsymbol{u} + ({\boldsymbol{\nabla}} \boldsymbol{u})^\top$. $\boldsymbol{\tau}_{NN}$ is the extra deviatoric stress due to viscosity differences (from an arbitrary constant viscosity $\eta_\infty$).

The boundary conditions on the velocity field, $\boldsymbol{u}$, are: the disturbance flow caused by the particle should diminish in the far-field,
\begin{equation}
    \boldsymbol{u} \rightarrow \boldsymbol{0} \qquad \text{as} \thickspace |\boldsymbol{r}| \rightarrow \infty,
\end{equation}
where $\boldsymbol{r} = \boldsymbol{x} - \boldsymbol{x}_c$, $\boldsymbol{x}_c$ is the center of the spheroid; and the fluid velocity  should satisfy no-slip conditions on the surface of the particle $S_p$,
\begin{equation}
    \boldsymbol{u}(\boldsymbol{x}\in S_p) = \boldsymbol{U} + \boldsymbol{\Omega} \times \boldsymbol{r} + \boldsymbol{u}^s.
\label{eqn:no_slip_BC}
\end{equation} 
The surface velocity, $\boldsymbol{u}^s$, arises from activity such as deformation or slip while the unknown translational and rotational velocities, $\boldsymbol{U}$ and $\boldsymbol{\Omega}$, are found by enforcing the dynamic conditions on the particle. 

We use the prolate spheroidal squirmer model to represent non-spherical active swimmers in this paper. This model is a reasonable representation of ciliates like \textit{Paramecium caudatum}, that utilize synchronized beating cilia to facilitate movement. The original spheroidal squirmer model developed by \citet{Keller1977} only includes one swimming mode, $\boldsymbol{u}^s = -B_1 (\boldsymbol{s} \cdot \boldsymbol{p}) \boldsymbol{s}$, where $\boldsymbol{s}$ is the unit tangent vector to the surface of the spheroidal microswimmer.  Subsequent studies have incorporated the contribution of a force-dipole into this model as a second mode. Following \citet{Theers2016} and \citet{Vangogh2022}, the slip velocity in our model is expressed as
\begin{align}
    \boldsymbol{u}^s & = -B_1 (\boldsymbol{s} \cdot \boldsymbol{p}) \boldsymbol{s} - B_2 \Big (\frac{\boldsymbol{r}}{a} \cdot \boldsymbol{p} \Big ) (\boldsymbol{s} \cdot \boldsymbol{p}) \boldsymbol{s}\,.
\end{align}
 The sign of squirming ratio $\beta = B_2 / B_1$ can be used to divide the swimmers into three types: pushers ($\beta < 0$), pullers ($\beta > 0$) and neutral swimmers ($\beta = 0$). Pushers, like \textit{E. coli}, generate propulsion from the back. \textit{Chlamydomonas reinhardtii}, on the other hand, is categorized as a puller because it uses its flagella to pull fluid from the front. Finally, neutral squirmers produce a flow corresponding to a source dipole. The two-mode spheroidal squirmer model simplifies to the spherical squirmer model in the case of zero eccentricity.

Recent research offers a more general representation of the flow field around a spheroidal squirmer, accounting for an infinite number of squirming modes \citep{Pohnl2020}. The swimming speed and stresslet of such a squirmer are influenced by more than just the $B_1$ and $B_2$ modes. However, these additional modes only significantly affect the outcome when the particle is notably slender \citep{Pohnl2020}, and so the two-mode prolate squirmer model is generally sufficient to depict swimming behavior \citep{Theers2016, Qi2020, Vangogh2022, Chi2022}. For simplicity we use only two modes in our calculations.

Finally, in the absence of inertia, the net force and torque on the particle must be zero
\begin{align}
\mathsf{\t F}_{ext}+\mathsf{\t F} = \boldsymbol{\mathsf{0}}
\end{align}
where $\mathsf{\t F} = [\boldsymbol{F} \thickspace \boldsymbol{L} ]^\top$ is a six-dimensional vector including both hydrodynamic force and torque, respectively
\begin{align}
\boldsymbol{F} &= \int_{S_p} \boldsymbol{n} \cdot \boldsymbol{\sigma} \thickspace \text{d} S, \label{eqn:BC3_a_f}\\
\boldsymbol{L} &= \int_{S_p} \boldsymbol{r} \times (\boldsymbol{n} \cdot \boldsymbol{\sigma}) \thickspace \text{d} S,
\label{eqn:BC3_a_t}
\end{align}
and $\boldsymbol{n}$ is the unit normal vector to the surface of the spheroidal particle. Whereas $\mathsf{\t F}_{ext} = [\boldsymbol{F}_{ext}\thickspace \boldsymbol{L}_{ext} ]^\top$ represents any external forces and torques acting on the particle. Enforcing this dynamic condition sets the particle's translational and rotational velocities.

\section{Reciprocal theorem \label{section:reciprocal_thm}}
Rather than solving the velocity field due to the spheroid directly, we instead use the reciprocal theorem to project onto operators from a known auxiliary flow in order to obtain the hydrodynamic force and torque. Following the approach outlined by \citet{Elfring2017}, active particle dynamics in a fluid of arbitrary rheology can be written as
\begin{equation}
       \mathsf{\t U} =  \hat{\mathsf{\t R}}^{-1}_{\mathsf{\t F} \mathsf{\t U}} \cdot ( \mathsf{\t F}_{ext} + \mathsf{\t F}_{s} + \mathsf{\t F}_{NN}),
       \label{eqn:RCvelocity}
\end{equation}
where $\mathsf{\t U} = [\boldsymbol{U} \thickspace \boldsymbol{\Omega} ]^\top$  is a six-dimensional vector including translational and rotational velocities. 

The term 
\begin{equation}
    \mathsf{\t F}_{s} = \int_{S_p} \boldsymbol{u}^s \cdot (\boldsymbol{n} \cdot \hat{ \mathsf{\t T}}_{ \mathsf{\t U}}) \thickspace \text{d} S,
    \label{eqn:propulsive_s}
\end{equation}
represents the propulsive force and torque exerted by the particle due to the slip velocity, $\boldsymbol{u}^s$, in a \textit{homogeneous} Newtonian fluid, while the term
\begin{equation}
    \mathsf{\t F}_{NN} = - \int_{\mathcal{V}} \boldsymbol{\tau}_{NN} : \hat{\mathsf{\t E}}_{\mathsf{\t U}} \thickspace \text{d} V,
    \label{eqn:NN}
\end{equation}
accounts for the additional force and torque stemming from the extra deviatoric stress, $\boldsymbol{\tau}_{NN}$, in the fluid volume $\mathcal{V}$ where the squirmer is immersed. 

The terms denoted with a hat are linear operators associated with the auxiliary flow solution of rigid-body motion of a body of the same shape in a homogeneous Newtonian fluid of viscosity $\eta_{\infty}$. The tensors $\hat{ \mathsf{\t T}}_{ \mathsf{\t U}}$ and $\hat{\mathsf{\t E}}_{\mathsf{\t U}}$ are spatially dependent functions that map velocities of the particle $\hat{ \mathsf{\t U}}$ to the stress $\hat{\boldsymbol{\sigma}} = \hat{\mathsf{\t T}}_{ \mathsf{\t U}} \cdot \hat{\mathsf{\t U}} $ and rate of strain $ \hat{\dot{\boldsymbol{\gamma}}} = 2 \hat{\mathsf{\t E}}_{\mathsf{\t U}} \cdot \hat{\mathsf{\t U}} $, respectively, while $\hat{\mathsf{\t R}}_{\mathsf{\t F} \mathsf{\t U}}$ is the ($6\times6$) resistance tensor. These operators are well known for prolate spheroids (see Appendix \ref{app:stokes} for further details).

The extra stress $\boldsymbol{\tau}_{NN}$ due to small viscosity variations, is parameterized by $\varepsilon$, so we expand all flow quantities in regular perturbation series in $\varepsilon$,
\begin{equation}
    \{ \boldsymbol{u}, \boldsymbol{\sigma}, \boldsymbol{\tau}_{NN}, \dot{\boldsymbol{\gamma}}, \boldsymbol{U}, \boldsymbol{\Omega}\} = \{ \boldsymbol{u}_0, \boldsymbol{\sigma}_0, \boldsymbol{0}, \dot{\boldsymbol{\gamma}}_0, \boldsymbol{U}_0, \boldsymbol{\Omega}_0 \} + \varepsilon \{ \boldsymbol{u}_1, \boldsymbol{\sigma}_1, \boldsymbol{\tau}_{NN,1}, \dot{\boldsymbol{\gamma}}_1, \boldsymbol{U}_1, \boldsymbol{\Omega}_1 \} + O (\varepsilon^2).
    \label{eqn:perturbation_analysis}
\end{equation}

At leading order, we have a homogeneous Newtonian fluid of viscosity $\eta_{\infty}$. Viscosity variations are captured at the next order, $O(\varepsilon)$, where the extra stress
\begin{align}
\boldsymbol{\tau}_{NN} = (\eta (\boldsymbol{x}) - \eta_{\infty}) \dot{\boldsymbol{\gamma}}_0 + O(\varepsilon^2),
\label{eq:tauNN}
\end{align}
and $\dot{\boldsymbol{\gamma}}_0$ is the strain rate of the flow of an active particle in the leading order homogeneous fluid. Upon substitution of \eqref{eq:tauNN} in \eqref{eqn:NN}, we see that calculation of the extra force and torque 
\begin{equation}
    \mathsf{\t F}_{NN} = - \int_{\mathcal{V}} (\eta (\boldsymbol{x}) - \eta_{\infty}) \dot{\boldsymbol{\gamma}}_0 : \hat{\mathsf{\t E}}_{\mathsf{\t U}} \thickspace \text{d} V+ O (\varepsilon^2),
    \label{eqn:FNN}
\end{equation}
due to spatial variations of viscosity, up to $O(\varepsilon)$, requires only the integration of known Stokes flow solutions, $\dot{\boldsymbol{\gamma}}_0$, and $\hat{\mathsf{\t E}}_{\mathsf{\t U}}$ from the auxiliary resistance problem. Analytical evaluation of the integral is most easily performed in a particle-aligned spheroidal coordinate system with details given in Appendix \ref{app:coordinates}.

It is important to note that when dealing with linearly varying viscosity fields such that $(\eta (\boldsymbol{x}) - \eta_{\infty}) \sim \varepsilon x$, the expansion maintains regularity only for $x \sim o(1/\varepsilon)$. However, the far-field contribution of a squirmer at distances $r \sim O(1/\varepsilon)$ is $O(\varepsilon^2)$ with respect to the non-Newtonian force $\boldsymbol{F}_{NN}$ and $O(\varepsilon^3)$ with respect to the non-Newtonian torque $\boldsymbol{L}_{NN}$. The velocity field of a passive spheroid decays slower than that of a squirmer; however, in constant viscosity gradients the far-field contribution to the integrals at $O(\varepsilon)$ is exactly zero (due to symmetry), making these systems suitable for analysis using a regular perturbation scheme.

\section{Passive spheroids\label{section:Passive}}
Before examining the dynamics of an active particle we first derive the mobility of a passive prolate spheroid subject to an external force and torque, $\mathsf{\t F}_{ext}$, in a viscosity gradient. For a passive spheroid there is no active slip $\boldsymbol{u}^s = \boldsymbol{0}$, and thus $\mathsf{\t F}_{s}=\boldsymbol{\mathsf{0}}$.

At leading order, $\mathsf{\t F}_{NN}=\boldsymbol{\mathsf{0}}$, and from \eqref{eqn:RCvelocity} we simply obtain the dynamics of a passive spheroid in a homogeneous Newtonian fluid of viscosity $\eta_{\infty}$, under an external force and torque $\mathsf{\t F}_{ext}$, which satisfies the usual mobility relationship \citep{Kim1991},
\begin{equation}
       \mathsf{\t U}_0 =  \hat{\mathsf{\t R}}^{-1}_{\mathsf{\t F} \mathsf{\t U}} \cdot  \mathsf{\t F}_{ext}.
       \label{eqn:passive_0th_order}
\end{equation}
The flow field at this order is identical to the auxiliary flow field in the previous section (Appendix \ref{app:stokes}) thus the strain rate, $\dot{\boldsymbol{\gamma}}_0 = 2 \hat{\mathsf{\t E}}_{\mathsf{\t U}} \cdot \mathsf{\t U}_0$, can be written as
\begin{equation}
    \dot{\boldsymbol{\gamma}}_0 = 2 \hat{\mathsf{\t E}}_{\mathsf{\t U}} \cdot \hat{\mathsf{\t R}}^{-1}_{\mathsf{\t F} \mathsf{\t U}}
   \cdot  \mathsf{\t F}_{ext}.
   \label{eqn:strain_rate_0th_order}
\end{equation}

At first order, substitution of \eqref{eqn:strain_rate_0th_order} into \eqref{eqn:FNN} yields
\begin{gather}
 \mathsf{\t F}_{NN}
 = -\boldsymbol{\mathsf{\t R}}_{NN}
   \cdot \hat{\mathsf{\t R}}^{-1}_{\mathsf{\t F} \mathsf{\t U}} \cdot \mathsf{\t F}_{ext},
   \label{eqn:passive_F_NN}
\end{gather}
where for convenience we have defined the tensor
\begin{align}
\boldsymbol{\mathsf{\t R}}_{NN} =\int_{\mathcal{V}}2(\eta (\boldsymbol{x}) - \eta_{\infty})\hat{\mathsf{\t E}}_{\mathsf{\t U}}:\hat{\mathsf{\t E}}_{\mathsf{\t U}}\, \text{d}V.
\label{eqn:R_NN}
\end{align}

Using \eqref{eqn:RCvelocity}, we obtain the translational and rotational velocity of a passive prolate spheroid at first order
\begin{gather}
\varepsilon\mathsf{\t U}_1
 =  
   -\hat{\mathsf{\t R}}^{-1}_{\mathsf{\t F} \mathsf{\t U}}
   \cdot \boldsymbol{\mathsf{\t R}}_{NN}
   \cdot \hat{\mathsf{\t R}}^{-1}_{\mathsf{\t F} \mathsf{\t U}} \cdot \mathsf{\t F}_{ext}.
   \label{eqn:passive_first_order}
\end{gather}

Combining \eqref{eqn:passive_0th_order} and \eqref{eqn:passive_first_order},
\begin{gather}
 \mathsf{\t U} = \mathsf{\t U}_0 + \varepsilon \mathsf{\t U}_1 = \left( \hat{\mathsf{\t R}}^{-1}_{\mathsf{\t F} \mathsf{\t U}}
   -\hat{\mathsf{\t R}}^{-1}_{\mathsf{\t F} \mathsf{\t U}}
   \cdot \boldsymbol{\mathsf{\t R}}_{NN}
   \cdot \hat{\mathsf{\t R}}^{-1}_{\mathsf{\t F} \mathsf{\t U}}\right) \cdot \mathsf{\t F}_{ext},
\end{gather}
we obtain the mobility $\mathsf{\t M}_{\mathsf{\t U} \mathsf{\t F}}  =  \hat{\mathsf{\t R}}^{-1}_{\mathsf{\t F}\mathsf{\t U}}-\hat{\mathsf{\t R}}^{-1}_{\mathsf{\t F} \mathsf{\t U}}
   \cdot \boldsymbol{\mathsf{\t R}}_{NN}
   \cdot \hat{\mathsf{\t R}}^{-1}_{\mathsf{\t F} \mathsf{\t U}}$, connecting the particle velocities $\mathsf{\t U}$ to the external force and torque $\mathsf{\t F}_{ext}$, valid to first order in $\varepsilon$, where 
\begin{gather}
\mathsf{\t M}_{\mathsf{\t U} \mathsf{\t F}}
   =
 \begin{pmatrix}
   \mathsf{\t M}_{\boldsymbol{U} \boldsymbol{F}} 
   \quad \mathsf{\t M}_{\boldsymbol{U} \boldsymbol{L}} \\
   \mathsf{\t M}_{\boldsymbol{\Omega} \boldsymbol{F}} 
   \quad \mathsf{\t M}_{\boldsymbol{\Omega} \boldsymbol{L}}
   \end{pmatrix},
   \label{eqn:mobility_matrix}
\end{gather}
and $\mathsf{\t M}_{\boldsymbol{U} \boldsymbol{L}} = \mathsf{\t M}_{\boldsymbol{\Omega} \boldsymbol{F}}^\top$. In homogeneous fluids, the mobility is determined solely by the shape and orientation of the particle, specified by the eccentricity $e$ and the orientation vector $\boldsymbol{p}$. In viscosity gradients, the mobility also depends on $\boldsymbol{\nabla} \eta$. The expressions for the force-translational velocity coupling, $\mathsf{\t M}_{\boldsymbol{U} \boldsymbol{F}}$, and the torque-angular velocity coupling, $\mathsf{\t M}_{\boldsymbol{\Omega} \boldsymbol{L}}$, are essentially identical to when the viscosity is constant, 
\begin{align}
    \mathsf{\t M}_{\boldsymbol{U} \boldsymbol{F}}  & = \frac{1}{6 \pi \eta |_{\boldsymbol{x}_c} a} [ \frac{1}{\mathcal{X}^{A}} \boldsymbol{p} \boldsymbol{p} + \frac{1}{\mathcal{Y}^{A}} (\mathsf{\t I} - \boldsymbol{p} \boldsymbol{p})], \\
    \mathsf{\t M}_{\boldsymbol{\Omega} \boldsymbol{L}} & = \frac{1}{8 \pi \eta |_{\boldsymbol{x}_c} a^3} [ \frac{1}{\mathcal{X}^{C}} \boldsymbol{p} \boldsymbol{p} + \frac{1}{\mathcal{Y}^{C}} (\mathsf{\t I} - \boldsymbol{p} \boldsymbol{p})].
\end{align}
except now the viscosity is now evaluated at the instantaneous particle center $\boldsymbol{x}_c$. The coefficients $\mathcal{X}^{A}$, $\mathcal{Y}^{A}$, $\mathcal{X}^{C}$, $\mathcal{Y}^{C}$ are functions of eccentricity $e$ and their expressions are given in Appendix \ref{app:stokes}.

Unlike in homogeneous Newtonian fluids, in viscosity gradients there arises a torque-translational velocity (and force-angular velocity) coupling
\begin{equation}
    \mathsf{\t M}_{\boldsymbol{U} \boldsymbol{L}}  = \frac{\varepsilon}{6 \pi \eta_{\infty} a^2}  \Big [ \Lambda_1 ( \boldsymbol{d} \times \mathsf{\t I}) + \Lambda_2 ( \boldsymbol{p} \cdot \boldsymbol{d}) ( \boldsymbol{p} \times \mathsf{\t I}) + \Lambda_3 (\boldsymbol{d} \times \boldsymbol{p}) \boldsymbol{p} \Big ],\label{eqn:B}
\end{equation}
where
\begin{align}
    \Lambda_1 & = \frac{3[- 2e + (1 - e^2) \mathcal{L}_e]}{16 e^3},   \\
    \Lambda_2 & = \frac{3[ 2e(-3 + e^2) + (3 - 2 e^2 + 2e^4) \mathcal{L}_e]}{32 e^3 (2 - e^2)},  \\
    \Lambda_3 & = \frac{3[ 2e(9 - 5 e^2) + (-9 + 8 e^2 + e^4) \mathcal{L}_e]}{32 e^3 (2 - e^2)},
    \label{eqn:Lambda_3}
\end{align}
and $\mathcal{L}_e =  \ln \left(\frac{1 + e}{1-e}\right)$. In the spherical limit ($e \rightarrow 0$), the inverse of the mobility in \eqref{eqn:mobility_matrix} agrees with the resistance tensor of a sphere reported previously by \citet{Datt2019}. 

We have also made a comparison between our calculations and the results for an elongated prolate spheroid sedimenting in viscosity gradients according to \citet{Kamal2023}. At large aspect ratios $\lambda = a/b \rightarrow \infty$ the mobilities can be written as
\begin{align}
   \mathsf{\t M}_{\boldsymbol{U} \boldsymbol{F}}  & \sim \frac{1}{6 \pi \eta |_{\boldsymbol{x}_c} a} [ \frac{3 \ln \lambda}{2} \boldsymbol{p} \boldsymbol{p} + \frac{3 \ln \lambda}{4} (\mathsf{\t I} - \boldsymbol{p} \boldsymbol{p})], \\
    \mathsf{\t M}_{\boldsymbol{U} \boldsymbol{L}}  & \sim \frac{\varepsilon}{6 \pi \eta_{\infty} a^2}  \Big [ - \frac{3}{8}  ( \boldsymbol{d} \times \mathsf{\t I}) + \frac{3 \ln \lambda}{4} ( \boldsymbol{p} \cdot \boldsymbol{d}) ( \boldsymbol{p} \times \mathsf{\t I}) + \frac{3}{4} (\boldsymbol{d} \times \boldsymbol{p}) \boldsymbol{p} \Big ], \\
    \mathsf{\t M}_{\boldsymbol{\Omega} \boldsymbol{L}} & \sim \frac{1}{8 \pi \eta |_{\boldsymbol{x}_c} a^3} [ \frac{3 \lambda^2}{2} \boldsymbol{p} \boldsymbol{p} + 3 \ln \lambda (\mathsf{\t I} - \boldsymbol{p} \boldsymbol{p})].
\end{align}
In this limit, we obtain the mobility matrix for an asymptotically slender spheroid in a constant viscosity gradient. Calculating the leading order translational and rotational velocities with external force $\boldsymbol{F}_{ext} = - m \boldsymbol{g}$ and torque $ \boldsymbol{L}_{ext} = \boldsymbol{0}$, our results exactly coincide with the sedimenting velocities of slender filaments in viscosity gradients found by \citet{Kamal2023}.

\begin{figure}
\centering
\includegraphics[scale = 0.5]{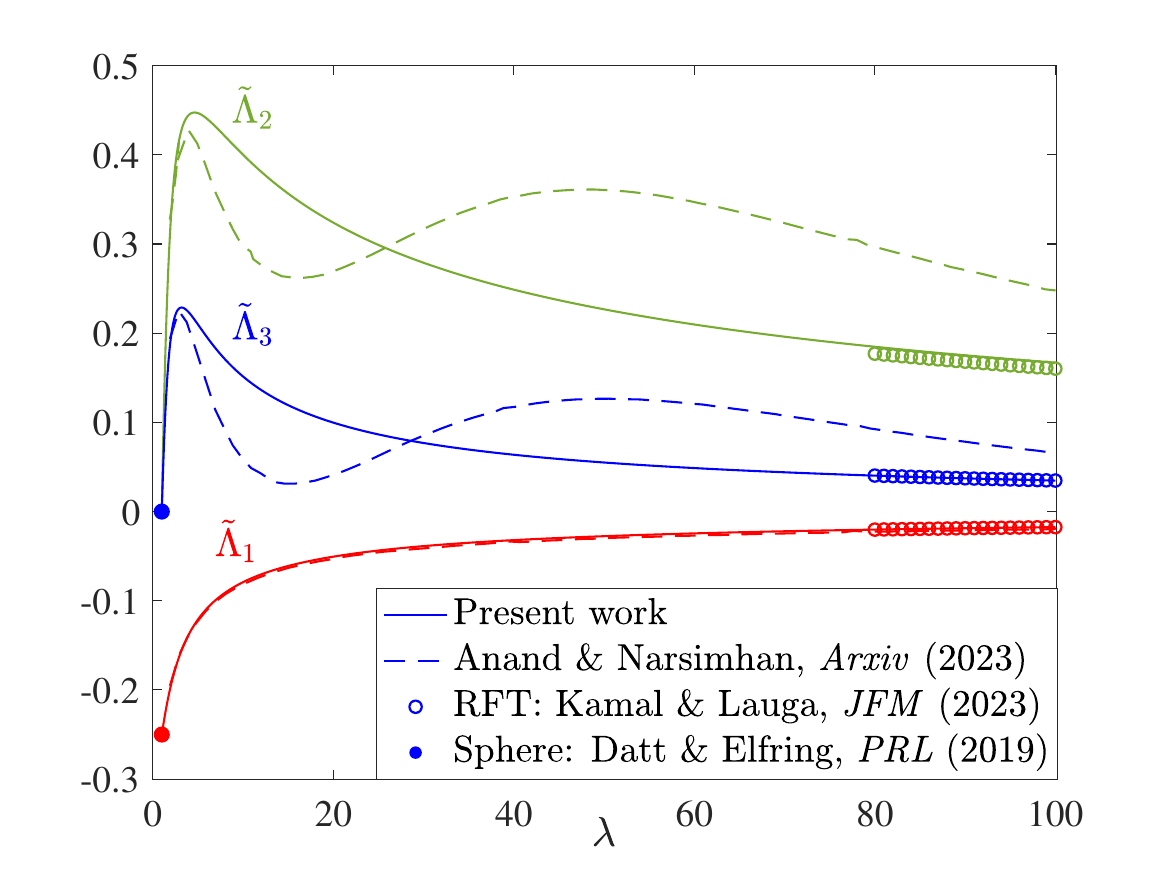}
\caption{\label{fig:comparison} A plot of mobility coefficients $\tilde{\Lambda_i}$ as a function of aspect ratio $\lambda$. Solid lines represent the present work, dashed lines are those found by \citet{Anand2023}. Also shown are the data for sphere from \citet{Datt2019} (filled symbols) and for an asymptotically slender spheroid from \citet{Kamal2023} (open symbols).}
\end{figure}

Recent work also explored the dynamics of sedimenting passive spheroids in viscosity gradients numerically \citep{Anand2023}. The authors of that work constructed a dimensionless mobility matrix and following their approach we rescale so that the dimensionless force-angular velocity tensor
\begin{equation}
    \tilde{\mathsf{\t M}}_{\boldsymbol{U} \boldsymbol{L}} = \frac{6 \pi \eta_{\infty} a^2}{\lambda^{4/3}}\mathsf{\t M}_{\boldsymbol{U} \boldsymbol{L}}  =  \tilde{\Lambda}_1 ( \boldsymbol{d} \times \mathsf{\t I}) + \tilde{\Lambda}_2 ( \boldsymbol{p} \cdot \boldsymbol{d}) ( \boldsymbol{p} \times \mathsf{\t I}) + \tilde{\Lambda}_3 (\boldsymbol{d} \times \boldsymbol{p}) \boldsymbol{p},
\end{equation}
where
\begin{equation}
    \tilde{\Lambda}_i = \frac{\varepsilon}{\lambda^{4/3}} \Lambda_i, \qquad i = 1,2,3.
\end{equation}
We then compare dimensionless coefficients $\tilde{\Lambda}_i$ with the corresponding numerical results by \citet{Anand2023} for different aspect ratios, as shown in Figure \ref{fig:comparison}. While agreement is very good with $\tilde{\Lambda}_1$, there is some discrepancy between our results for $\tilde{\Lambda}_2$ and $\tilde{\Lambda}_3$, and those found by \citet{Anand2023}. A possible reason for the discrepancies may be due to the difference in the definition of perturbation parameter. Unlike our perturbation parameter $\varepsilon$ which dictates the viscosity variations across the particle, $\Delta \eta / \eta_{\infty}$, are always small, \citet{Anand2023} use the perturbation parameter $ \varepsilon \lambda^{-2/3}$, which for a fixed small value can lead to large viscosity differences near the particle at large $\lambda$. As another point of comparison, we also calculate the corresponding values of $\tilde{\Lambda}_i$ from \cite{Kamal2023} and, as shown in Figure \ref{fig:comparison}, when the aspect ratio is large, our analytical results align closely.

\section{Active Spheroids\label{section:Active}}
Microswimmers are often considered to be neutrally buoyant, we do the same here and hence we assume there is no externally applied force or torque, $\mathsf{\t F}_{ext}=\boldsymbol{\mathsf{0}}$, on an the active spheroid swimming in a viscosity gradient. 

At leading order in $\varepsilon$, we have an active spheroid swimming in a homogeneous Newtonian fluid of viscosity $\eta_{\infty}$. The swim speed is well known \citep{Keller1977,Theers2016,Pohnl2020,Vangogh2022},
\begin{gather}
 \mathsf{\t U}_0
 =
   \hat{\mathsf{\t R}}^{-1}_{\mathsf{\t F} \mathsf{\t U}}
   \cdot 
\mathsf{\t F}_s
=
    \begin{pmatrix}
    \frac{2e - (1 - e^2) \mathcal{L}_e}{2e^3} B_1 \boldsymbol{p}  \\
    \boldsymbol{0}
   \end{pmatrix}.
   \label{eqn:zeroth_order}
\end{gather}
The corresponding flow field is given in Appendix \ref{app:stokes}.

At first order, the translational and rotational velocities, 
\begin{gather}
\varepsilon\mathsf{\t U}_1
 =
   \hat{\mathsf{\t R}}^{-1}_{\mathsf{\t F} \mathsf{\t U}}
   \cdot 
\mathsf{\t F}_{NN},
   \label{eqn:first_order}
\end{gather}
are obtained using \eqref{eqn:FNN}, with $\dot{\boldsymbol{\gamma}}_0$ calculated from the flow field solutions of a two-mode active spheroidal squirmer in Appendix \ref{app:stokes}. Combining \eqref{eqn:zeroth_order} with \eqref{eqn:first_order}, $\mathsf{\t U} = \mathsf{\t U}_0 + \varepsilon \mathsf{\t U}_1$, we obtain expressions valid up to $O(\varepsilon)$ for the translational and rotational velocities of a prolate spheroidal squirmer
\begin{align}
    \label{eqn:tran_vel}
    \boldsymbol{U} & = \boldsymbol{U}_0 - \frac{a B_2}{5} (\mathcal{X}^{U} \boldsymbol{I} - \mathcal{Y}^{U} 3 \boldsymbol{pp}) \cdot {\boldsymbol{\nabla}} \left(\frac{\eta }{ \eta_{\infty}}\right), \\
    \boldsymbol{\Omega} & = - \frac{1}{2} \mathcal{X}^{\Omega} \boldsymbol{U}_{0} \times {\boldsymbol{\nabla}} \left(\frac{\eta }{ \eta_{\infty}}\right),
    \label{eqn:rot_vel}
\end{align}
where the coefficients
\begin{align}
    \mathcal{X}^U & = \frac{5 \left[ -6e + 4 e^3 + 3 \left(1 - e^2\right) \mathcal{L}_e \right] \left[ - 6e + 10 e^3 + 3 \left(1 - e^2\right)^2 \mathcal{L}_e\right]}{24 e^5 \left[ 6e - \left(3 - e^2\right)\mathcal{L}_e\right]}, \\
    \mathcal{Y}^U & = \frac{5 \left[ -6e + 4 e^3 + 3 \left(1 - e^2\right) \mathcal{L}_e\right] \left[ - 18e + 6 e^3 + \left(9 - 6e^2 + 5e^4\right) \mathcal{L}_e\right]}{72 e^5 \left[ 6e - \left(3 - e^2\right) \mathcal{L}_e\right]}, \\
    \mathcal{X}^{\Omega} & = \frac{\left(1 - e^2\right) \left[-2 e + \left(1 + e^2\right) \mathcal{L}_e \right]}{\left(2 - e^2\right) \left[ 2e - \left(1 - e^2\right) \mathcal{L}_e \right]},
\end{align}
are monotonically decreasing functions of the eccentricity. In the spherical limit, $e\rightarrow 0$, $\mathcal{X}^U=1$, $\mathcal{Y}^U=1$ and $\mathcal{X}^{\Omega}=1$ and we exactly recover the dynamics for spheres found by \citet{Datt2019}. Conversely in the slender limit, $e\rightarrow 1$, $\mathcal{X}^U=0$, $\mathcal{Y}^U=5/9$ and $\mathcal{X}^{\Omega}=0$, meaning infinitely slender squirmers do not reorient in viscosity gradients, but there is still a change in their translational velocity due to the interaction of the dipolar flow with the spatial variations in viscosity. Generally (for $e \lessapprox 0.9988$), the speed change is greater for spheroids than spheres when aligned with the gradient. 

In general, the behavior of spheroidal squirmers is qualitatively similar to spherical squirmers as they navigate through constant viscosity gradients \citep{Datt2019}: all swimmers display negative viscotaxis by reorienting to swim down viscosity gradients, except that the impact of the gradient is diminished with increasing slenderness. The mechanistic reason for this change is straight forward, the viscosity difference across a slimmer body is reduced, which leads to slower reorientation and in the slender limit viscotaxis ceases.

\begin{figure}
\centering 
\subfloat{\includegraphics[scale = 0.35]{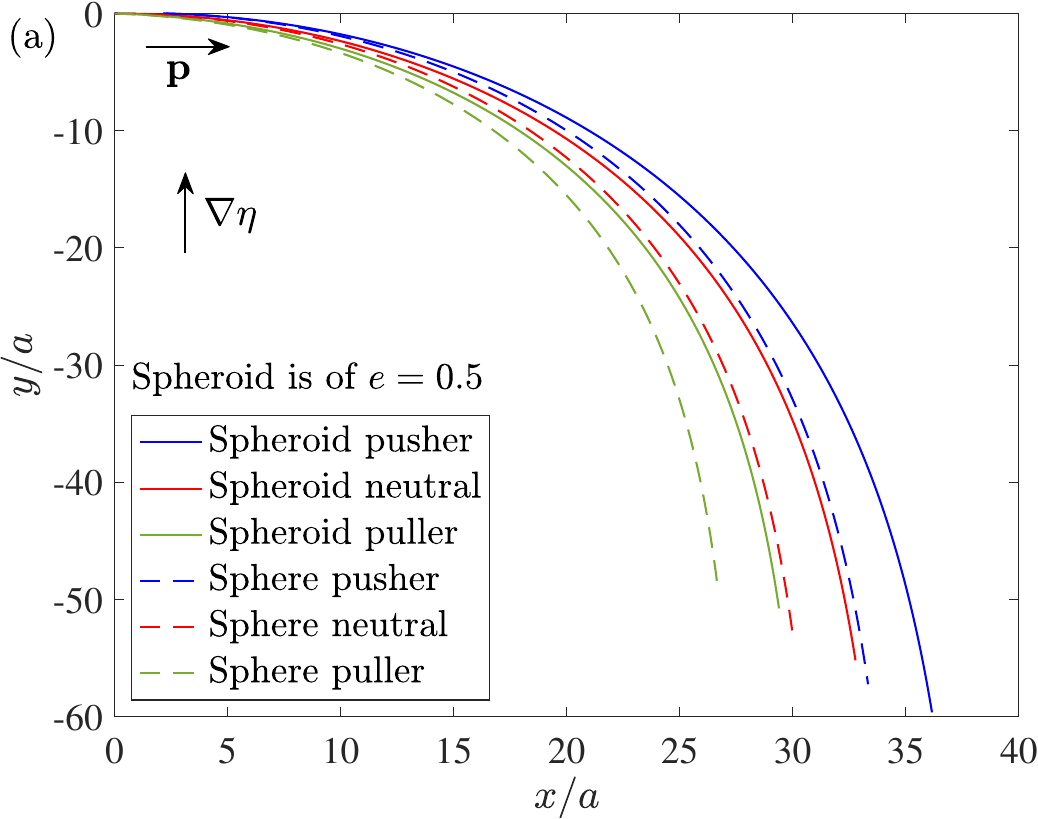}\label{fig:figure3a}}
\subfloat{\includegraphics[scale = 0.35]{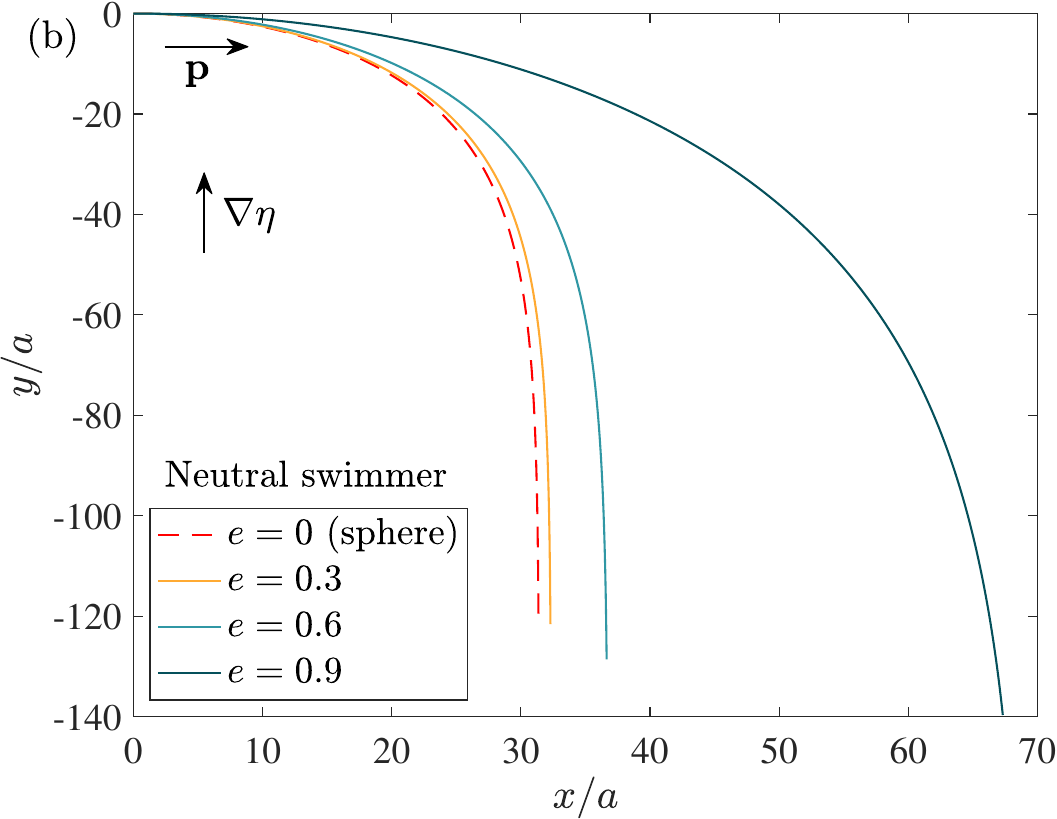}\label{fig:figure3b}}\hfill
\caption{(a) Trajectories of spheroidal $(e = 0.5)$ and spherical squirmers with an initial orientation $\boldsymbol{p}$ orthogonal to the viscosity gradient $\boldsymbol{\nabla} \eta$ between $t=0$ to $t = 100a/B_1$. (b) Trajectories of neutral spheroidal squirmers of different eccentricities swimming at an initial orientation $\boldsymbol{p}$ orthogonal to the viscosity gradient $\boldsymbol{\nabla} \eta$ from $t=0$ to $t = 250a/B_1$. All squirmers eventually swim down the viscosity gradient.}
\end{figure}

In Fig.~\ref{fig:figure3a}, we compare trajectories of spherical squirmers and spheroidal squirmers ($e = 0.5$) for all three types of swimmers ($\beta = \pm 2$ for pullers and pushers and $\varepsilon = 0.1$). Spheroidal pushers still exhibit the greatest range of movement, traversing both horizontally across the gradient and vertically along it, whereas pullers cover the least distance. As expected,  Fig.~\ref{fig:figure3a} shows that spheroidal squirmers take longer to reorient than spherical squirmers. In Fig.~\ref{fig:figure3b} we show the effect on a neutral squirmer as the eccentricity increases, making the spheroid more elliptical in shape, illustrating that the effect on the dynamics becomes dramatic for increasingly slender swimmers.

\begin{figure}
\centering 
\subfloat{\includegraphics[scale = 0.24]{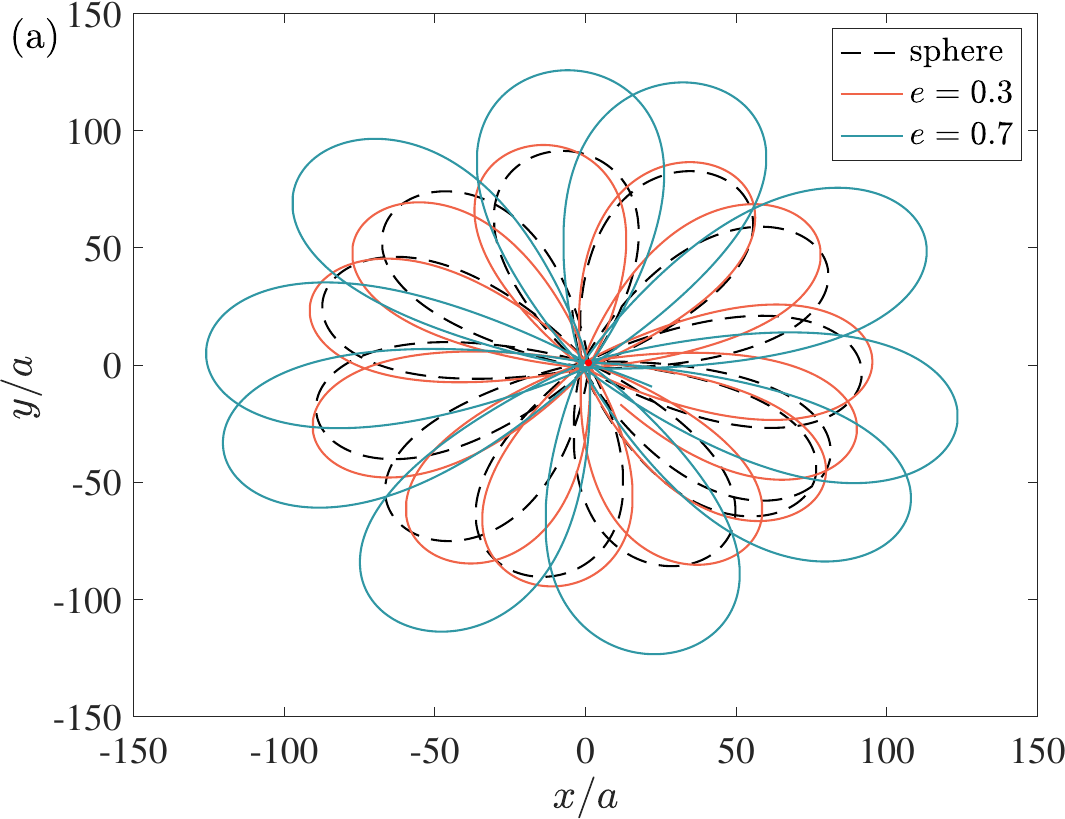}\label{fig:figure4a}}
\subfloat{\includegraphics[scale = 0.24]{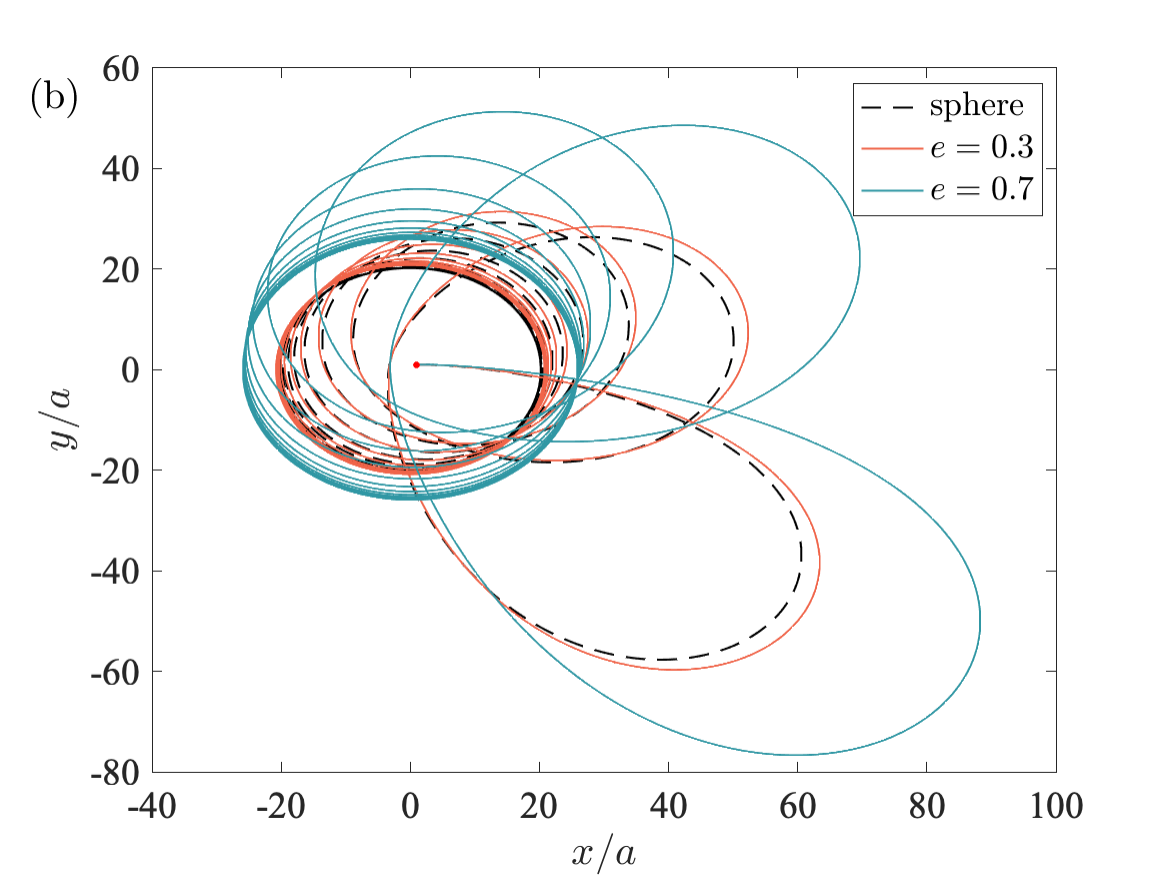}\label{fig:figure4b}}
\subfloat{\includegraphics[scale = 0.24]{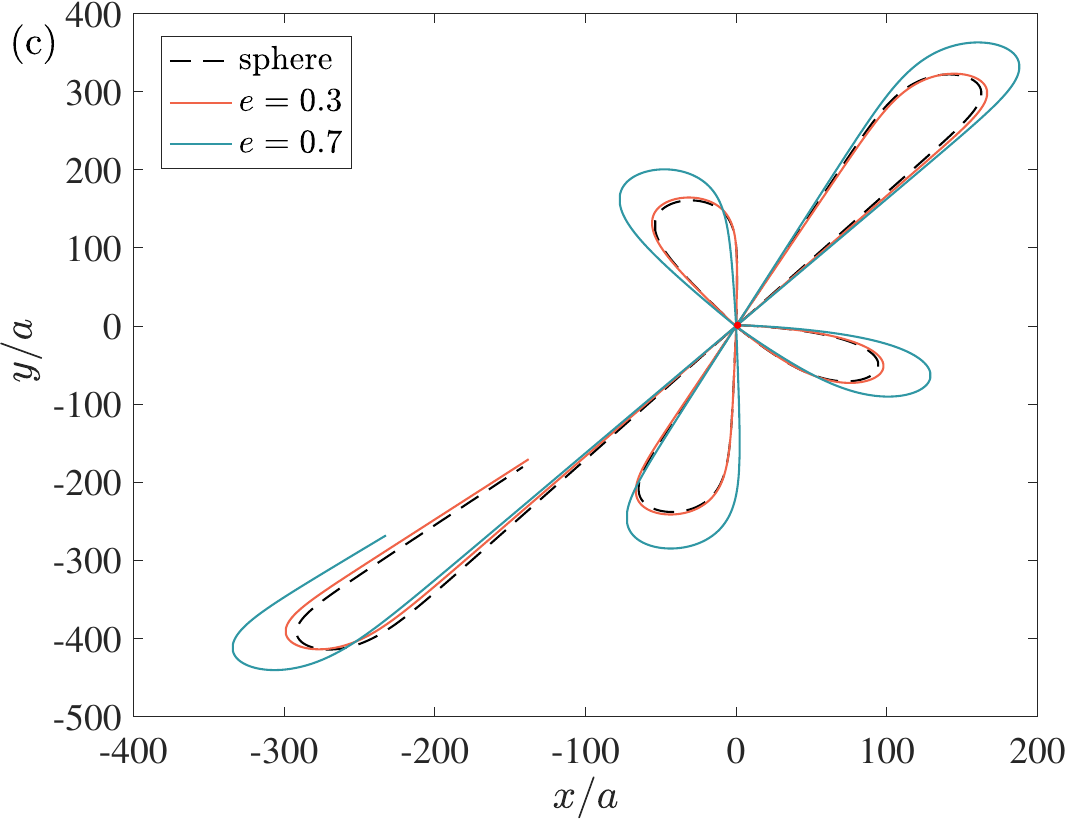}\label{fig:figure4c}}
\caption{\label{fig:traj} Planar trajectories of three types of spheroidal swimmers are depicted, (a) neutral swimmers, (b) pushers, and (c) pullers. The initial position of each swimmer is ($x/a = 1$, $y/a = 1$), indicated by a red dot, with the swimmers initially pointing in the positive $x$-axis direction. These swimmers are placed in a radial viscosity gradient, where the viscosity increases radially outward from the original point. The dynamics of the spheroidal squirmers qualitatively resembles that of spherical swimmers, except that the reorientation is slowed and so orbits have a larger radius.}
\end{figure}

We also plot, in Fig.~\ref{fig:traj}, the trajectories of squirmers swimming in a radially varying viscosity field
\begin{align}
\boldsymbol{\nabla} (\eta/\eta_{\infty}) = \varepsilon \boldsymbol{e}_r /a,
\end{align}
as shown by \citet{Datt2019} for spheres. Here the assumption is that equations ~\eqref{eqn:tran_vel} and ~\eqref{eqn:rot_vel} still hold as a \textit{local} approximation of dynamics even in radial viscosity gradients because, at the particle length scale, the distinctions between the two types of gradients should be minimal. In this viscosity field, the dynamics of all three types of spheroidal squirmers again closely resemble those of spherical squirmers except that the reorientation dynamics is slowed as the squirmers become more slender. In particular, as with spheres, pushers and neutral swimmers have a stable orbit about the viscosity minimum and as the particle becomes more slender, the radius of that orbit correspondingly expands.

\section{Disturbance viscosity effects \label{section:Disturbance}}
Up to this point we have assumed spatial variations in viscosity are prescribed and not disturbed by the presence of the particle. However,  because variations in the viscosity generally arise from variations in an underlying field that affects the viscosity such as temperature, salt, or nutrient concentration, we should take into account the effect of boundary conditions on the surface of the particle for that underlying field. For example, in an otherwise linear salt concentration field, the presence of a particle may disrupt the field (and thus the coupled viscosity field) due to salt impermeability, or in an otherwise linear temperature field the particle may disrupt the field due to differences in thermal conductivity between the fluid and the particle. Although these disturbances diminish with distance from the particle, the disturbance does have a leading order effect on the dynamics of the active particle \citep{Shaik2021}. 

Here, we determine the dynamics of a prolate spheroid swimmer in an otherwise constant viscosity gradient while considering the disturbance viscosity caused by a no-flux condition on the boundary of the particle following the work of \citet{Shaik2021} for spheres. We write total viscosity field as the superposition of an ambient viscosity field (denoted as $\eta_0$) and a disturbance viscosity field (denoted by prime),
\begin{align}
    \eta = \eta_0 + \eta',
\end{align}
where the disturbance viscosity diminishes in the far-field region
\begin{equation}
    \eta' \rightarrow 0 \qquad \text{as} \thickspace |\boldsymbol{r}| \rightarrow \infty. \label{eqn:BC_disturbance_viscosity}
\end{equation}

The transport of a scalar like temperature or salt concentration is governed by an advection-diffusion equation. When the scalar variations are weak, the changes in viscosity are directly proportional to the changes in the underlying scalar field, hence, viscosity transport is governed by a similar advection-diffusion equation. For microswimmers moving slowly in a highly diffusive scalar such as temperature or salt concentration, advection is usually small. In this limit, the distribution of viscosity satisfies Laplace's equation. As the ambient viscosity field is linear, the disturbance viscosity must also satisfy Laplace's equation,
\begin{equation}
     \nabla^2 \eta = \nabla^2 \eta' = 0. \label{eqn:disturbance_viscosity}
\end{equation}

The disturbance viscosity is also determined by the boundary conditions present on the particle's surface. Here we consider that the surface is impermeable to nutrient or salt concentration, or insulating to the temperature. In this scenario, the particle surface maintains a no-flux condition for viscosity, where
\begin{align}
\boldsymbol{n} \cdot \boldsymbol{\nabla} \eta = 0 \quad \text{on} \quad S_p.
\end{align}
The detailed disturbance viscosity field is given in Appendix \ref{app:disturbance} (where we also give solutions with an alternative boundary condition $\eta(\boldsymbol{x}\in S_p)=const$). Here we only explain the effect of disturbance viscosity on the dynamics of the active spheroid.

The impact of the total viscosity field (both ambient and disturbance viscosities) on the swimming velocity of a particle with a no-flux condition is, to leading order
\begin{align}
    \boldsymbol{U}_1 & = - \frac{13 a B_2}{60} (\mathcal{X}^{U,nf} \mathsf{\t I} - \mathcal{Y}^{U,nf} 3 \boldsymbol{pp}) \cdot \boldsymbol{\nabla} \left(\frac{\eta_0 }{ \eta_{\infty}}\right) \label{eqn:no_flux_translational_velocity}, \\
    \boldsymbol{\Omega}_1 & = - \frac{5}{8} \mathcal{X}^{\Omega,nf} \boldsymbol{U}_{0} \times {\boldsymbol{\nabla}} \left(\frac{\eta_0 }{ \eta_{\infty}}\right), \label{eqn:no_flux_rotational_velocity}
\end{align}
where
\begin{align}
    \mathcal{X}^{U,nf} & = 5 \Big [ 4e^2(63 - 117e^2 + 52 e^4 ) + 12 e (1 - e^2)^2 (-21 + 2e^2) \mathcal{L}_e  - 9 (-7 + e^2) (1 - e^2)^2  \mathcal{L}_e^2 - 6e  (1 - e^2)^2 \mathcal{L}_e^3 \Big ] \nonumber \\
    & \qquad \times \Big \{ 13 e^2 [6e + (-3+e^2)\mathcal{L}_e][-2e+4e^3-(-1+e^2)\mathcal{L}_e] \Big \}^{-1}, \\
    \mathcal{Y}^{U,nf} & = 10 \Big [ 8e^4(48 - 77e^2 + 32 e^4 ) - 4 e^3 \left( 75 - 129e^2 + 58 e^4\right) \mathcal{L}_e \nonumber \\
    & \quad + 2e^4 ( 33 - 56e^2 + 23e^4 ) \mathcal{L}_e^2 - e (1 - e^2)^2 (-33 + 37e^2)  \mathcal{L}_e^3 - 3(1 - e^2)^3 \mathcal{L}_e^4 \Big ] \nonumber \\
    & \qquad \times \Big \{ 39 e [6e + (-3+e^2)\mathcal{L}_e][-2e+4e^3-(-1+e^2)\mathcal{L}_e][2e+(-1+e^2)\mathcal{L}_e] \Big \}^{-1}, \\
    \mathcal{X}^{\Omega,nf} & = \frac{2 \left(1 - e^2\right) \left[ 4 e^2 (7 - 8e^2) + 4e(-7 + 7e^2 + 2e^4) \mathcal{L}_e - (-7 + 6e^2 + e^4) \mathcal{L}_e^2\right]}{5\left(2 - e^2\right) \left[ -2e + 4e^3 +\left(1 - e^2\right) \mathcal{L}_e \right] [2e + (-1+e^2)\mathcal{L}_e]},
\end{align}
are monotonically decreasing functions of the eccentricity. In the spherical limit, $e\rightarrow 0$, $\mathcal{X}^U=1$, $\mathcal{Y}^U=1$ and $\mathcal{X}^{\Omega}=1$ and we exactly recover the dynamics for spheres found by \citet{Shaik2021}. Conversely in the slender limit, $e\rightarrow 1$, $\mathcal{X}^U=0$, $\mathcal{Y}^U=20/39$ and $\mathcal{X}^{\Omega}=0$.

We see that the disturbance viscosity does not alter the fundamental physics of a spheroidal particle governed in comparison to effects of the ambient viscosity alone. It primarily increases the rate at which the particle rotates to align against the viscosity gradient. It also enhances the effects of the ambient viscosity field on various swimmer types: pushers speed up, pullers slow down, while neutral swimmers maintain consistent speeds relative to those in a homogeneous fluid.

\section{Conclusion \label{section:Conclusion}}

In this paper, we analyzed the hydrodynamics of prolate spheroids, both passive and active, in constant viscosity gradients. For passive spheroids, we determined the mobility tensor that governs the dynamics of a spheroid under an external force and torque in viscosity gradients. Our analytical expression agrees with, and generalizes, previous results for spheres \citep{Datt2019} and asymptotically slender bodies \citep{Kamal2023}. We also derived formulas for the dynamics of active spheroids in constant viscosity gradients. These results generalize previous results for active spherical squirmers \citep{Datt2019, Shaik2021} to include the effects of particle shape. In general, the behavior of spheroidal squirmers is qualitatively similar to spherical squirmers as they navigate through constant viscosity gradients. All swimmers display negative viscotaxis by reorienting to swim down viscosity gradients, except that the impact of the gradient is diminished with increasing slenderness. The viscosity difference across their body is reduced for slimmer swimmers, which leads to slower reorientation and in the slender limit viscotaxis ceases. The implications of this may seem limited but it actually raises interesting new possibilities. For example, consider a swimmer that consists of a slim `tail' that produces thrust but is too slender to drive reorientation in a viscosity gradient, coupled with a large spherical passive `head' that strongly interacts with a viscosity gradient. Our results (for passive and active bodies) indicate that such a swimmer would display \textit{positive} viscotaxis by reorienting to swim up viscosity gradients in a fashion analogous to what was originally proposed by \citet{Liebchen2018}. Extending this idea further, one can see that geometry and activity can be tailored to control or eliminate viscotaxis. These results enrich the current understanding of how particle shape impacts viscotaxis, and the insights gleaned from this study may have implications not only for understanding the complex dynamics of natural microswimmers, but also for guiding the design and manipulation of synthetic active particles in complex fluidic systems.\\

\noindent\textbf{Funding.} This work was supported by the Natural Sciences and Engineering Research Council of Canada (RGPIN-2020-04850) and by a UBC Killam Accelerator Research Fellowship to G.J.E.\\


\appendix

\section{Spheroids in Stokes flow}\label{app:stokes}
Here we give solutions to the Stokes equations for a passive and active spheroid in a Newtonian fluid with constant viscosity.

\subsection{Spheroidal multipoles}
Before proceeding to the solution of a passive prolate spheroid we first introduce the spheroidal multipole solutions \citep{Chwang1975a} that are used. 

The Green's function, $\mathsf{\t G}$, of the Stokes equations and derivatives are, in component form
 \begin{align}
G_{ij}&=\frac{\delta_{ij}}{r} +\frac{x_ix_j}{r^3}, & \text{Stokeslet}, \nonumber\\
G^d_{ijk}&=G_{ij,k}=\frac{\delta_{jk}x_i+\delta_{ik}x_j-\delta_{ij}x_k}{r^3} -3\frac{x_ix_j}{r^5}, & \text{dipole},\nonumber\\
G^D_{ij}&=G_{ij,ll}=2\frac{\delta_{ij}}{r^3} -6\frac{x_ix_j}{r^5}, & \text{potential doublet},\nonumber\\
G^R_{ijk}&=\frac{1}{2} (G_{ij,k}-G_{ik,j})=\frac{\delta_{ik}x_j-\delta_{ij}x_k}{r^3}, & \text{rotlet},\nonumber\\
G^{S}_{ijk}&=\frac{1}{2} (G_{ij,k}+G_{ik,j})=\frac{\delta_{kj}x_i}{r^3}-3\frac{x_ix_jx_k}{r^5}, & \text{stresslet},\nonumber\\
G^Q_{ijk}&=G_{ij,llk}=-6\frac{\delta_{jk}x_i+\delta_{ik}x_j+\delta_{ij}x_k}{r^5} +30 \frac{x_ix_jx_k}{r^7}, & \text{potential quadrupole}.\nonumber
\end{align}
Spheroidal multipoles are a weighted distribution of the above multipoles between the foci $\xi =-c$ to $c$, where $c = ae$, used to represent flows around spheroidal particles,
\begingroup
\allowdisplaybreaks
\begin{align}
Q_{ij}&=\int^{c}_{-c}  G_{ij}(\boldsymbol{x}-\xi \boldsymbol{p}) d\xi, \nonumber\\
Q^D_{ij}&=\int^{c}_{-c} (c^2-\xi^2)G^D_{ij}(\boldsymbol{x}-\xi \boldsymbol{p}) d\xi ,\nonumber\\
Q^R_{ijk}&=\int^{c}_{-c}  (c^2-\xi^2)G^R_{ijk}(\boldsymbol{x}-\xi \boldsymbol{p}) d\xi,\nonumber\\
Q^{S}_{ijk}&=\int^{c}_{-c}(c^2-\xi^2)G^{S}_{ijk}(\boldsymbol{x}-\xi \boldsymbol{p})  d\xi ,\nonumber\\
Q^Q_{ijk}&=\int^{c}_{-c}  (c^2-\xi^2)^2G^Q_{ijk}(\boldsymbol{x}-\xi \boldsymbol{p}) d\xi. \nonumber
\end{align}
\endgroup

Explicit expressions for spheroidal multipoles are taken from \citet{Einarsson2015} and \citet{Abtahi2019}
\begingroup
\allowdisplaybreaks
\begin{align}
Q_{ij}&=\delta_{ij} I^0_1+x_ix_j  I^0_3 -(x_ip_j+x_jp_i) I^1_3+ p_ip_jI^2_3,\nonumber\\
Q^D_{ij}&=2\delta_{ij}J^0_3+6 \Big[-x_ix_jJ^0_5+(x_ip_j+x_jp_i)J_5^1-p_ip_jJ^2_5\Big],\nonumber\\
Q^R_{ijk}&=(\delta_{ik}x_j-\delta_{ij}x_k)J^0_3+(\delta_{ij}p_k-\delta_{ik}p_j)J^1_3,\nonumber\\
Q^{S}_{ijk}&=\delta_{jk}x_iJ^0_3-\delta_{jk}p_iJ^1_3   \nonumber\\
                        &\quad+3 \Big[ -x_ix_jx_kJ^0_5+(x_ix_kp_j+x_jx_kp_i+x_ix_jp_k)J^1_5\nonumber\\
                &\hspace{1cm}-(x_kp_ip_j+x_ip_jp_k+x_jp_ip_k)J^2_5+p_ip_jp_kJ^3_5\Big],\nonumber\\
Q^Q_{ijk}&=6 \Big [-(\delta_{jk}x_i+\delta_{ik}x_j+\delta_{ij}x_k)K^0_5+(\delta_{jk}p_i+\delta_{ik}p_j+\delta_{ij}p_k)K^1_5 \Big]\nonumber\\
                       &\quad+30 \Big[ x_ix_jx_kK^0_7-(x_ix_kp_j+x_jx_kp_i+x_ix_jp_k)K^1_7 \nonumber\\
                &\hspace{1cm}+(x_kp_ip_j+x_ip_jp_k+x_jp_ip_k)K^2_7-p_ip_jp_kK^3_7 \Big],\nonumber\\
Q_{ij,k} & = (- \delta_{ij} x_k +  \delta_{ik} x_j + \delta_{jk} x_i) I^0_3 + (\delta_{ij} p_k -  \delta_{ik} p_j - \delta_{jk} p_i) I^1_3 \nonumber \\
        & \quad + 3 \Big[ - x_i x_j x_k I^0_5 + (x_ix_kp_j+x_jx_kp_i+x_ix_jp_k) I^1_5 \Big ], \nonumber \\
Q^D_{ij,k} & = 6 \Big[ - ( \delta_{ij} x_k +  \delta_{ik} x_j + \delta_{jk} x_i) J^0_5 + ( \delta_{ij} p_k +  \delta_{ik} p_j + \delta_{jk} p_i) J^1_5 \Big ] \nonumber \\
        & \quad + 30 \Big[ x_i x_j x_k J^0_7 - (x_ix_kp_j+x_jx_kp_i+x_ix_jp_k) J^1_7 \nonumber \\
        & \quad + (x_kp_ip_j+x_ip_jp_k+x_jp_ip_k) J^1_7 - p_ip_jp_k J^3_7 \Big ], \nonumber \\
Q^{R}_{ijk,m}&=(\delta_{ik}\delta_{jm}-\delta_{ij}\delta_{km})J^0_3 \nonumber\\
        & \quad+3(\delta_{ik}x_j-\delta_{ij}x_k)( p_m J^1_5-x_m J^0_5)\nonumber\\
          &  \quad+3(\delta_{ij}p_k-\delta_{ik}p_j)(p_m J^2_5-x_m J^1_5),\nonumber\\
Q^{S}_{ijk,m}&=\delta_{jk}\delta_{im}J^0_3+3\delta_{jk}x_i (p_mJ^1_5-x_mJ^0_5) -3\delta_{jk}p_i(p_mJ^2_5-x_mJ^1_5) \nonumber\\
                        &\quad+3 \Big[ -(\delta_{im}x_jx_k+\delta_{jm}x_ix_k+\delta_{km}x_ix_j)J^0_5-5x_ix_jx_k(p_mJ^1_7-x_mJ^0_7) \nonumber\\
                        &\quad+(\delta_{im}x_kp_j+\delta_{km}x_ip_j  + \delta_{jm} x_kp_i+\delta_{km} x_jp_i    +  \delta_{im} x_jp_k+\delta_{jm}x_ip_k)J^1_5 \nonumber\\
                         &\quad+5(x_ix_kp_j+x_jx_kp_i+x_ix_jp_k)(p_mJ^2_7-x_mJ^1_7) \nonumber\\
                        &\quad-(\delta_{km}p_ip_j+\delta_{im}p_jp_k+\delta_{jm}p_ip_k)J^2_5 +5p_ip_jp_k(p_mJ^4_7-x_mJ^3_7 \nonumber\\
                &\quad-5(x_kp_ip_j+x_ip_jp_k+x_jp_ip_k)(p_mJ^3_7-x_mJ^2_7)\Big],\nonumber\\ 
Q^Q_{ijk,m}&=6 \Big [-(\delta_{jk}\delta_{im}+\delta_{ik}\delta_{jm}+\delta_{ij}\delta_{km})K^0_5-5(\delta_{jk}x_i+\delta_{ik}x_j+\delta_{ij}x_k)(p_mK^1_7-x_mK^0_7)\nonumber\\
                      &\quad+5(\delta_{jk}p_i+\delta_{ik}p_j+\delta_{ij}p_k)(p_mK^2_7-x_mK^1_7) \Big]\nonumber\\
                      &\quad+30 \Big[ (\delta_{im}x_jx_k+\delta_{jm}x_ix_k+\delta_{km}x_ix_j)K^0_7+7x_ix_jx_k(p_mK^1_9-x_mK^0_9)\nonumber\\
                     &\quad-(\delta_{im}x_kp_j+\delta_{km}x_ip_j+\delta_{jm}x_kp_i+\delta_{km}x_jp_i+\delta_{im}x_jp_k+\delta_{jm}x_ip_k)K^1_7\nonumber\\
                       &\quad-7(x_ix_kp_j+x_jx_kp_i+x_ix_jp_k)(p_mK^2_9-x_mK^1_9)\nonumber\\
                     &\quad+(\delta_{km}p_ip_j+\delta_{im}p_jp_k+\delta_{jm}p_ip_k)K^2_7 -p_ip_jp_kK^3_7\nonumber\\
             &\quad+7(x_kp_ip_j+x_ip_jp_k+x_jp_ip_k)(p_mK^3_9-x_mK^2_9) \Big],\nonumber
\end{align}
\endgroup
where
 \begin{align}
 \label{integI}
I^n_m &=\int^{c}_{-c} d\xi \frac{\xi^n}{|\boldsymbol{x}-\xi \boldsymbol{p}|^m},  \\   
J^n_m &=c^2 I^n_m -I^{n+2}_m, \label{integJ}\\ 
K^n_m&=c^2 J^n_m -J^{n+2}_m=c^4 I^n_m - 2 c^2I^{n+2}_m +I^{n+4}_m.
\label{integK}
\end{align}

The integrals $I^n_{m}$ satisfy the relationship
 \begin{align}
&\frac{\partial}{\partial x_i} I^n_{m} = mp_i  I^{n+1}_{m+2} - mx_i I^{n}_{m+2}.
 \end{align}
To simplify integration one may employ an auxiliary coordinate system, $(x',y',z')$, with $x'$ aligned with $\boldsymbol{p}$ such that
\begin{align}
& I^n_m=\int^c_{-c} d\xi \frac{\xi^n}{[(x'-\xi)^2+(y')^2+(z')^2]^{m/2}}=\int^c_{-c} d\xi \frac{\xi^n}{[(x'-\xi)^2+R^2]^{m/2}},
 \end{align}
where on the surface of the particle we have
 \begin{align}
& R_1=\sqrt{\left(x'+c\right)^2+R^2},\nonumber\\
& R_2=\sqrt{\left(x'-c\right)^2+R^2},\nonumber\\
& R=\sqrt{\left(1-e^2\right) \left(a^2-x'^2\right)}.
 \end{align}
The integrals also satisfy the relationship
 \begin{align}
&I^n_{m} =  x' I^{n-1}_m+\frac{(n-1) I^{n-2}_{m-2}}{m-2}-\frac{c^{n-1}
   \left((-1)^{n} R_1^{2-m}+R_2^{2-m}\right)}{m-2}.
 \end{align}
Integrals $J^n_m$, and $K^n_m$ can be calculated easily from equations \eqref{integJ} and \eqref{integK}. 

\subsection{A passive prolate spheroid}
\subsubsection{Rigid-body translation}
The flow field due to a prolate spheroid translating with velocity $\hat{\boldsymbol{U}}$ in a quiescent fluid 
  \begin{align}
  \label{eqn:rigid_body_translation}
\hat{u}_{i}= (Q_{ij} + \alpha_1 Q^D_{ij}) \Big [\mathcal{A}^U p_j p_m + \mathcal{B}^U (\delta_{jm}-p_jp_m) \Big ] \hat{U}_{m},
 \end{align}
where
 \begin{align}
 &\alpha_1=\frac{1-e^2}{4e^2},\nonumber\\
&\mathcal{A}^U=\frac{e^2}{ - 2e +  \left(1 + e^2\right)\mathcal{L}_e},\nonumber\\
&\mathcal{B}^U=\frac{2e^2}{ 2e +  \left(3e^2 - 1 \right)\mathcal{L}_e}.
 \end{align}

The strain-rate tensor can be written
   \begin{align}
\hat{\dot{\boldsymbol{\gamma}}} &= 2 \hat{\boldsymbol{E}}_{\boldsymbol{U}} \cdot \hat{\boldsymbol{U}},
\end{align}
where
\begin{align} 
\hat{E}_{U_{ikm}} =\frac{1}{2} (Q^T_{ijk} + \alpha_1 Q^{DT}_{ijk}) \Big [\mathcal{A}^U p_j p_m + \mathcal{B}^U (\delta_{jm}-p_jp_m) \Big ],
 \end{align}
and $Q^{T}_{ijk}=Q_{ij,k}+Q_{kj,i}$. $Q^{DT}_{ijk}$ is defined similarly to $Q^T_{ijk}$.

\subsubsection{Rigid-body rotation}
The flow field due to a prolate spheroid rotating with angular velocity $\hat{\boldsymbol{\Omega}}$ in an other quiescent fluid is
  \begin{align}
  \label{eqn:rigid_body_rotation}
\hat{u}_{i}=&\Big \{-\epsilon_{jkl} Q^R_{ijk} \big[ \mathcal{A}^{\Omega} p_l p_s +\mathcal{B}^{\Omega} (\delta_{ls}-p_lp_s)\big ] \nonumber\\
&+\big( Q^{S}_{ijk}+ \alpha_2 Q^Q_{ijk} \big)  \mathcal{C}^{\Omega} (\epsilon_{jsm}p_kp_m+\epsilon_{ksm}p_jp_m) \Big \} \hat{\Omega}_{s},
 \end{align}
where
\begingroup
 \begin{align}
 &\alpha_2=\frac{1-e^2}{8e^2},\nonumber\\
&\mathcal{A}^{\Omega}=\frac{1 - e^2}{-4 e + 2 \left(1 - e^2\right) \mathcal{L}_e},\nonumber\\
&\mathcal{B}^{\Omega}=\frac{2 - e^2}{4 e - 2 \left(1 + e^2\right) \mathcal{L}_e},\nonumber\\
&\mathcal{C}^{\Omega}=\frac{e^2}{ 4e - 2 \left(1 + e^2\right) \mathcal{L}_e}.
\end{align}
 \endgroup

The strain-rate tensor can be written
   \begin{align}
\hat{\dot{\boldsymbol{\gamma}}} = 2 \hat{\boldsymbol{E}}_{\boldsymbol{\Omega}} \cdot \hat{\boldsymbol{\Omega}},
\end{align}
where
\begin{align} 
\hat{E}_{\Omega_{ims}} =\frac{1}{2} \Big \{ & -\epsilon_{jkl} Q^{RT}_{ijkm} \big [\mathcal{A}^{\Omega} p_l p_s +\mathcal{B}^{\Omega} (\delta_{ls}-p_lp_s) \big ] +\big( Q^{ST}_{ijkm}+ \alpha Q^{QT}_{ijkm} \big) \mathcal{C}^{\Omega} (\epsilon_{jsm}p_kp_m+\epsilon_{ksm}p_jp_m)\Big \},
 \end{align}
and $Q^{RT}_{ijkm}=Q^{R}_{ijk,m}+Q^{R}_{mjk,i}$.  $Q^{ST}_{ijkm}$ and $Q^{QT}_{ijkm}$ are defined similarly as $Q^{RT}_{ijkm}$.

Finally the tensor $\hat{\mathsf{\t E}}_{\mathsf{\t U}}$ used in the integral \eqref{eqn:NN} is simply defined as
\begin{align}
\hat{\mathsf{\t E}}_{\mathsf{\t U}} =
\begin{pmatrix}
\hat{\boldsymbol{E}}_{\boldsymbol{U}}\\
\hat{\boldsymbol{E}}_{\boldsymbol{\Omega}}
\end{pmatrix}.
\end{align}

\subsubsection{Mobility tensor for a prolate spheroid in a Newtonian fluid}
The mobility tensor,
\begin{gather}
 \hat{\boldsymbol{\mathsf{\t M}}}_{\mathsf{\t U} \mathsf{\t F}}
 = \hat{\mathsf{\t R}}^{-1}_{\mathsf{\t F} \mathsf{\t U}} = 
 \begin{pmatrix}
   \hat{\mathsf{\t M}}_{\boldsymbol{U} \boldsymbol{F}} 
   \quad \hat{\mathsf{\t M}}_{\boldsymbol{U} \boldsymbol{L}} \\
   \hat{\mathsf{\t M}}_{\boldsymbol{\Omega} \boldsymbol{F}} 
   \quad \hat{\mathsf{\t M}}_{\boldsymbol{\Omega} \boldsymbol{L}}
   \end{pmatrix},
   \label{eqn:newtonian_resistance_matrix}
\end{gather}
couples force and torque to rigid-body translation and rotation for a body in Stokes flows. For a prolate spheroid in a Newtonian fluid with constant viscosity $\eta_{\infty}$ there is no torque-translation (or force-rotation) coupling. Specifically the terms are \citep{Kim1991}
\begin{align}
     \hat{\mathsf{\t M}}_{\boldsymbol{U} \boldsymbol{F}}  &  =  \frac{1}{6 \pi \eta_{\infty} a} [\frac{1}{\mathcal{X}^{A}} \boldsymbol{pp} + \frac{1}{\mathcal{Y}^{A}} (\mathsf{\t I} - \boldsymbol{pp})], \nonumber \\
     \hat{\mathsf{\t M}}_{\boldsymbol{\Omega} \boldsymbol{L}} &  =  \frac{1}{8 \pi \eta_{\infty} a^3} [\frac{1}{\mathcal{X}^{C}} \boldsymbol{pp} + \frac{1}{\mathcal{Y}^{C}} (\mathsf{\t I} - \boldsymbol{pp})], \nonumber \\
     \hat{\mathsf{\t M}}_{\boldsymbol{U} \boldsymbol{L}} &  = \hat{\mathsf{\t M}}_{\boldsymbol{\Omega} \boldsymbol{F}} = \mathsf{\t 0},
\end{align}
 where $\mathcal{X}^{A}$, $\mathcal{Y}^{A}$, $\mathcal{X}^{C}$, and $\mathcal{Y}^{C}$ are functions of eccentricity $e$
\begin{align}
    &\mathcal{X}^A=\frac{8 e^3}{3[ - 2e +  \left(1 + e^2\right)\mathcal{L}_e]},\nonumber\\
    &\mathcal{Y}^A=\frac{16 e^3}{3[ 2e +  \left(3 e^2 - 1\right)\mathcal{L}_e]},\nonumber\\
    &\mathcal{X}^C=\frac{4e^3(1 - e^2)}{ 2e -  \left( 1 - e^2\right)\mathcal{L}_e},\nonumber\\
    &\mathcal{Y}^C=\frac{4e^3(2 - e^2)}{ -2e +  \left( 1 + e^2\right)\mathcal{L}_e}.
\end{align}

\subsubsection{Extra stress tensor}
In \eqref{eqn:R_NN} we defined the tensor
\begin{align*}
\boldsymbol{\mathsf{\t R}}_{NN} =\int_{\mathcal{V}}2(\eta (\boldsymbol{x}) - \eta_{\infty})\hat{\mathsf{\t E}}_{\mathsf{\t U}}:\hat{\mathsf{\t E}}_{\mathsf{\t U}}\, \text{d}V.
\end{align*}
Writing
\begin{gather}
 \boldsymbol{\mathsf{\t R}}_{NN} = 
 \begin{pmatrix}
   \mathsf{\t R}_{\boldsymbol{F} \boldsymbol{U}} 
   \quad \mathsf{\t R}_{\boldsymbol{F} \boldsymbol{\Omega}} \\
   \mathsf{\t R}_{\boldsymbol{L} \boldsymbol{U}} 
   \quad \mathsf{\t R}_{\boldsymbol{L} \boldsymbol{\Omega}}
   \end{pmatrix},
\end{gather}
we have
\begin{align}
    \mathsf{\t R}_{\boldsymbol{F} \boldsymbol{U}} = \int_{\mathcal{V}} 2 (\eta (\boldsymbol{x}) - \eta_{\infty}) \hat{\boldsymbol{E}}_{\boldsymbol{U}} : \hat{\boldsymbol{E}}_{\boldsymbol{U}} \thickspace \text{d} V, \nonumber \\
    \mathsf{\t R}_{\boldsymbol{F} \boldsymbol{\Omega}} = \int_{\mathcal{V}} 2 (\eta (\boldsymbol{x}) - \eta_{\infty})  \hat{\boldsymbol{E}}_{\boldsymbol{U}} : \hat{\boldsymbol{E}}_{\boldsymbol{\Omega}} \thickspace \text{d} V, \nonumber \\
    \mathsf{\t R}_{\boldsymbol{L} \boldsymbol{U}} = \int_{\mathcal{V}} 2 (\eta (\boldsymbol{x}) - \eta_{\infty}) \hat{\boldsymbol{E}}_{\boldsymbol{\Omega}} : \hat{\boldsymbol{E}}_{\boldsymbol{U}} \thickspace \text{d} V, \nonumber \\
    \mathsf{\t R}_{\boldsymbol{L} \boldsymbol{\Omega}} = \int_{\mathcal{V}} 2 (\eta (\boldsymbol{x}) - \eta_{\infty}) \hat{\boldsymbol{E}}_{\boldsymbol{\Omega}} : \hat{\boldsymbol{E}}_{\boldsymbol{\Omega}} \thickspace \text{d} V,
\end{align}
and $\mathsf{\t R}_{\boldsymbol{F} \boldsymbol{\Omega}} = \mathsf{\t R}_{\boldsymbol{L} \boldsymbol{U}} ^\top$.

\subsection{An active prolate spheroid in Stokes flow}
  The flow field $\boldsymbol{u}_0$ around a two-mode spheroidal squirmer swimming in a Newtonian fluid with constant viscosity  can written in terms of a stream function $\psi_0$ \citep{Keller1977,Theers2016,Vangogh2022}, using a particle-aligned coordinate system $O'\zeta_1 \zeta_2 \phi$ (see Appendix \ref{app:coordinates} for details), as 
  \begin{equation}
      \boldsymbol{u}_0 = \frac{1}{h_{\zeta_2} h_{\phi}} \frac{\partial \psi_0}{\partial \zeta_2} \boldsymbol{e}_{\zeta_1} - \frac{1}{h_{\zeta_1} h_{\phi}} \frac{\partial \psi_0}{\partial \zeta_1} \boldsymbol{e}_{\zeta_2},
      \label{active_spheroid_flow_field}
  \end{equation}
where
 \begin{align}
     \psi_0 & = C_1 H_2 (\zeta_1) G_2 (\zeta_2) + C_2 \zeta_1 ( 1 - \zeta_2^2)  + C_3 H_3 (\zeta_1) G_3 (\zeta_2) + C_4 \zeta_2 ( 1 - \zeta_2^2) + \frac{1}{2} U_0 c^2 (\zeta_1^2 - 1)(1 - \zeta_2^2).
 \end{align}
 Here $H_n(x)$ and $G_n(x)$ are Gegenbauer functions of the first and second order of degree $-1/2$ \citep{Theers2016}. The coefficients $C_n$ are
 \begin{align}
     C_1 & = 2c^2 \frac{U_0 (\tilde{\zeta_1}^2 + 1) - 2 B_1 \tilde{\zeta_1}^2}{- \tilde{\zeta_1} + (1 + \tilde{\zeta_1}^2 ) \coth^{-1} \tilde{\zeta_1}}, \nonumber \\
     C_2 & = c^2 \frac{B_1  \tilde{\zeta_1} [\tilde{\zeta_1} - (\tilde{\zeta_1}^2 - 1) \coth^{-1} \tilde{\zeta_1}) - U_0]}{- \tilde{\zeta_1} + (1 + \tilde{\zeta_1}^2) \coth^{-1} \tilde{\zeta_1}}, \nonumber \\
     C_3 & = c^2 \frac{ 4 B_2 \tilde{\zeta_1}}{3 \tilde{\zeta_1} + (1 - 3\tilde{\zeta_1}^2) \coth^{-1} \tilde{\zeta_1}}, \nonumber \\
     C_4 & = c^2 \frac{ B_2 \tilde{\zeta_1} [2/3 - \tilde{\zeta_1}^2 + \tilde{\zeta_1}(\tilde{\zeta_1}^2 - 1) \coth^{-1} \tilde{\zeta_1}]}{3 \tilde{\zeta_1} + (1 - 3\tilde{\zeta_1}^2) \coth^{-1} \tilde{\zeta_1}}
 \end{align}
and $\tilde{\zeta_1} = 1 / e$.

\section{Coordinate transformation}\label{app:coordinates}
We choose an arbitrary point $O$ and construct a lab-frame Cartesian coordinate system with unit vectors $\boldsymbol{e}_i$ ($i = \{1, 2, 3\}$) and position vector $\boldsymbol{x} = x\boldsymbol{e}_1 + y\boldsymbol{e}_2 + z\boldsymbol{e}_3$. The center of the particle can be expressed as $\boldsymbol{x}_c = x_c\boldsymbol{e}_1 + y_c\boldsymbol{e}_2 + z_c\boldsymbol{e}_3$. Without loss of generality, we can always adjust the axes to make sure that the ambient viscosity only varies in the $\boldsymbol{e}_1$ direction. However, the volume integrations in the reciprocal theorem are difficult to evaluate analytically in the lab-frame coordinate system $Oxyz$. To solve this problem, we use a particle-aligned Cartesian coordinate $O'XYZ$ and the corresponding spheroidal coordinate system $O'\zeta_1 \zeta_2 \phi$, where $O'$ is the center of the spheroid at $\boldsymbol{x}_c$. The Cartesian coordinate axes are determined by the viscosity gradient direction $\boldsymbol{d} = \frac{{\boldsymbol{\nabla}} \eta}{|{\boldsymbol{\nabla}}  \eta|}$ and the swimming direction $\boldsymbol{p}$. The unit vectors
\begin{align}
    \boldsymbol{e}_X & = \frac{(\boldsymbol{d} \times \boldsymbol{p}) \times \boldsymbol{p}}{|(\boldsymbol{d} \times \boldsymbol{p}) \times \boldsymbol{p}|}, \nonumber \\
    \boldsymbol{e}_Y & = \frac{\boldsymbol{d} \times \boldsymbol{p}}{|\boldsymbol{d} \times \boldsymbol{p}|}, \label{eqn: particle_aligned_coordinate}\\
    \boldsymbol{e}_Z & = \boldsymbol{p}. \nonumber
\end{align}
When $\boldsymbol{d}$ is parallel or anti-parallel to $\boldsymbol{p}$, we can simply set $\boldsymbol{e}_X = \boldsymbol{e}_1$, $\boldsymbol{e}_Y = \boldsymbol{e}_2$ and $\boldsymbol{e}_Z = \boldsymbol{e}_3$ without loss of generality. The position vector in this coordinate system $\boldsymbol{r} = \boldsymbol{x} - \boldsymbol{x}_c = X \boldsymbol{e}_X + Y \boldsymbol{e}_Y + Z \boldsymbol{e}_Z$. We then write the viscosity field
\begin{equation}
    \eta = \eta_{\infty} + \varepsilon \frac{\eta_{\infty}}{a} (x_c + \boldsymbol{r} \cdot \boldsymbol{e}_1),
\end{equation}
where $\boldsymbol{e}_1$ is obtained by inverting Eq.\eqref{eqn: particle_aligned_coordinate}.

In the particle-aligned Cartesian coordinate system, the surface of a spheroid satisfies
\begin{equation}
    \frac{Z^2}{a^2} + \frac{X^2 + Y^2}{b^2} = 1.
\end{equation}
Cartesian coordinates $(X, Y, Z)$ can be written in terms of $(\zeta_1,  \zeta_2, \phi)$  as
\begin{align}
    X & = c \sqrt{\zeta_1^2 - 1} \sqrt{ 1 - \zeta_2^2} \cos \phi, \nonumber \\
    Y & = c \sqrt{\zeta_1^2 - 1} \sqrt{ 1 - \zeta_2^2} \sin \phi,  \\
    Z & = c \zeta_1 \zeta_2, \nonumber
\end{align}
where $ 1 \leqslant \zeta_1 < \infty$, $ -1 \leqslant \zeta_2 \leqslant 1$ and $0 \leqslant \phi < 2\pi $. $c = \sqrt{a^2 - b^2}$ is half of the focal length. The unit normal vector and the tangent vector to the particle are $\boldsymbol{n} = \boldsymbol{e}_{\zeta_1}$ and $\boldsymbol{s} = -\boldsymbol{e}_{\zeta_2}$. Scale factors are
\begin{align}
    h_{\zeta_1} &= c \sqrt{\frac{\zeta_1^2 - \zeta_2^2}{\zeta_1^2 - 1}},\nonumber\\
     h_{\zeta_2} &= c \sqrt{\frac{\zeta_1^2 - \zeta_2^2}{1 - \zeta_2^2}},\\ 
     h_{\phi} &= c \sqrt{\zeta_1^2 - 1} \sqrt{ 1 - \zeta_2^2}.\nonumber
\end{align}

\begin{figure}
\centering 
\includegraphics[scale = 0.42]{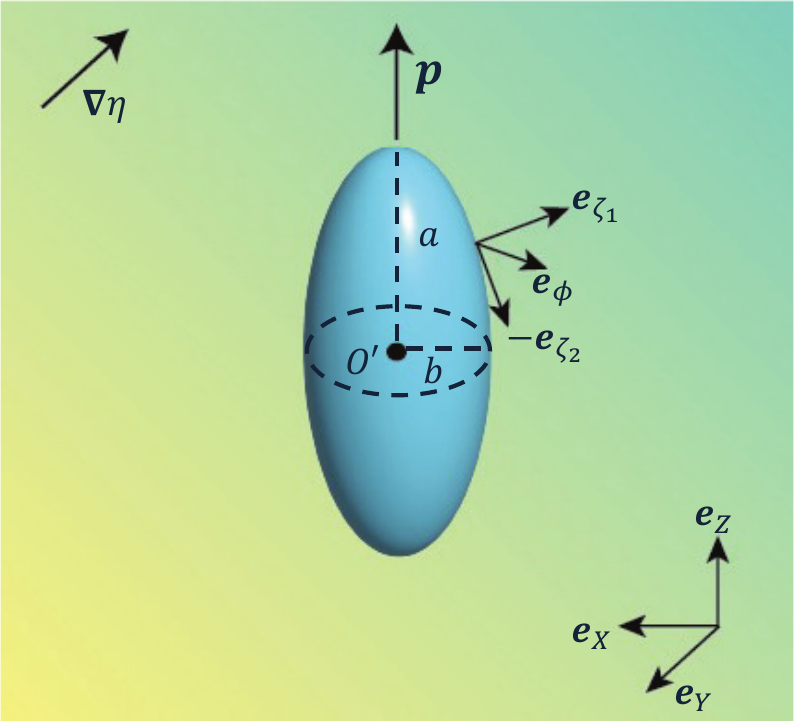}
\caption{\label{fig:schematic2} The particle-aligned cartesian coordinate system $O'XYZ$ and prolate spheroid coordinate $O'\zeta_1 \zeta_2 \phi$.} 
\end{figure}

\section{Disturbance viscosity field}\label{app:disturbance}
The general solution of \eqref{eqn:disturbance_viscosity} in spheroidal coordinates satisfies
\begin{equation}
    \eta '  = \sum_{k = 0}^{ \infty} \sum_{m=k}^{\infty} [A_{k,m} \cos (m \phi) + B_{k,m} \sin (m \phi)] P_k^m (\zeta_2) Q_k^m (\zeta_1),
\end{equation}
where $A_{k,m}$, $B_{k,m}$ are the constant coefficients, while $P_k^m$ and $Q_k^m$ are the associated Legendre polynomial of the first and second kind, respectively $k$ is the degree and $m$ is the order. Mathematical expressions of $P_k^m$ and $Q_k^m$ can be found in \citet{Abramowitz1964}. Below we determine the coefficients first for a no-flux boundary and then a constant viscosity boundary condition.

\subsection{No flux}
Supposing the ambient viscosity field is aligned with $\boldsymbol{e}_1$, we can write the no-flux constraint in particle-aligned coordinates as
\begin{equation}
    \frac{\partial \eta '}{\partial \zeta_1} \Big |_{\zeta_1 = \tilde{\zeta_1}} = - \varepsilon \eta_{\infty} p_1 e \zeta_2 + \frac{\varepsilon \eta_{\infty} \sqrt{1 - p_1^2}}{e \sqrt{1 - e^2}} \sqrt{1 - \zeta_2^2} \cos \phi,
\end{equation}
where $p_1 = \boldsymbol{p} \cdot \boldsymbol{e}_1$. The expression of disturbance viscosity field is
\begin{equation} 
    \eta ' = A_{1,0}  P_1^0 (\zeta_2) Q_1^0 (\zeta_1) + A_{1,1}  P_1^1 (\zeta_2) Q_1^1 (\zeta_1) \cos (\phi),
\end{equation}
where
\begin{align}
     A_{1,0} & = \varepsilon \eta_{\infty} p_1 \frac{2e (1 - e^2)}{ \left[ 2e - \left(1 - e^2\right) \mathcal{L}_e\right]}, \nonumber \\
     A_{1,1} & = \varepsilon \eta_{\infty} \sqrt{1 - p_1^2} \frac{2e (1 - e^2)}{ \left[ 2e - 4 e^3 - \left(1 - e^2\right) \mathcal{L}_e\right]}. \nonumber 
\end{align}

The changes in the translational and rotational velocity due to disturbance viscosity are
\begin{align}
    \boldsymbol{U}_1' & = - \frac{a B_2}{60} (\mathcal{X}^{U',nf} \boldsymbol{I} - \mathcal{Y}^{U',nf} 3 \boldsymbol{pp}) \cdot {\boldsymbol{\nabla}} (\frac{\eta_0 }{ \eta_{\infty}}), \label{eqn:nf_disturbance_tran}\\
    \boldsymbol{\Omega}_1' & = - \frac{1}{8} \mathcal{X}^{\Omega',nf} \boldsymbol{U}_{0} \times {\boldsymbol{\nabla}} (\frac{\eta_0 }{ \eta_{\infty}}), \label{eqn:nf_disturbance_ang}
\end{align}
where
\begin{align*}
    \mathcal{X}^{U',nf} & = 5 (-1 + e^2) \Big [ 8e^3(-9 -33e^2 + 32 e^4 ) - 4 e^2 \left( - 27 - 39e^2 + 62e^4\right) \mathcal{L}_e \\
    & \qquad + 6e ( -9 + 7e^2 -3e^4 + 5e^6) \mathcal{L}_e^2 - 3 \left(-3 + 9e^2 - 13e^4 +7 e^6\right) \mathcal{L}_e^3 \Big ] \\
    & \qquad \times \Big \{ 2 e^5 [6e + (-3+e^2)\mathcal{L}_e][-2e+4e^3-(-1+e^2)\mathcal{L}_e] \Big \}^{-1}, \\
    \mathcal{Y}^{U',nf} & = 5 (-1 + e^2) \Big [ 16e^4(-27 + 54e^2 - 102 e^4 + 64 e^6 ) - 16 e^3 \left( - 54 + 117e^2 -135 e^4 +74e^6 \right) \mathcal{L}_e \\
    & \quad + 8e^2 ( -81 + 189e^2 - 150e^4 + 31e^6 + 13 e^8) \mathcal{L}_e^2 - 4e \left(-54 + 135e^2 - 99e^4 +  e^6 + 17 e^8\right) \mathcal{L}_e^3 - 3(3-4e^2+e^4)^2 \mathcal{L}_e^4 \Big ] \\
    & \qquad \times \Big \{ 6 e^5 [6e + (-3+e^2)\mathcal{L}_e][-2e+4e^3-(-1+e^2)\mathcal{L}_e][2e+(-1+e^2)\mathcal{L}_e] \Big \}^{-1}, \\
    \mathcal{X}^{\Omega',nf} & = \frac{2 \left(1 - e^2\right) \left[ 4 e^2 (5 - 4e^2) + 20e(-1+e^2) \mathcal{L}_e + (5 - 6e + e^4) \mathcal{L}_e^2\right]}{\left(2 - e^2\right) \left[ -2e + 4e^3  + \left(1 - e^2\right) \mathcal{L}_e \right] [2e + (-1+e^2)\mathcal{L}_e]}.
\end{align*}
Adding \eqref{eqn:nf_disturbance_tran} and \eqref{eqn:nf_disturbance_ang} to \eqref{eqn:tran_vel} and \eqref{eqn:rot_vel} one obtains \eqref{eqn:no_flux_translational_velocity} and \eqref{eqn:no_flux_rotational_velocity}.

\subsection{Constant Viscosity}
The constant viscosity constraint, $\eta (x \in S_p) = \eta_p = const$, in spheroidal coordinates is
\begin{equation}
    \eta' \Big |_{\zeta_1 = \tilde{\zeta_1}} = \eta_c - \varepsilon \eta_{\infty} p_1  \zeta_2 + \varepsilon \eta_{\infty} \sqrt{(1 - p_1^2)(1 - e^2)(1 - \zeta_2^2)} \cos \phi,
\end{equation}
where $\eta_c = \eta_p - \eta_{\infty} - \varepsilon \frac{\eta_{\infty}}{a} x_c$, is a constant, while the other term varies on the surface of the spheroid. The disturbance viscosity field satisfying this constraint is
\begin{equation} 
    \eta '  = \acute{A}_{0,0} P_0^0 (\zeta_2) Q_0^0 (\zeta_1) +  \acute{A}_{1,0}  P_1^0 (\zeta_2) Q_1^0 (\zeta_1) + \acute{A}_{1,1}  P_1^1 (\zeta_2) Q_1^1 (\zeta_1) \cos (\phi), \label{eqn:constant_viscosity_field}
\end{equation}
where
\begin{align}
     \acute{A}_{0,0} & = \frac{2 \eta_c}{\mathcal{L}_e}, \nonumber \\
     \acute{A}_{1,0} & = \varepsilon \eta_{\infty} p_1 \frac{2e}{ 2e - \mathcal{L}_e}, \nonumber \\
     \acute{A}_{1,1} & = \varepsilon \eta_{\infty} \sqrt{1 - p_1^2} \frac{2e (1 - e^2)}{ \left[ 2e - \left(1 - e^2\right) \mathcal{L}_e\right]}.\nonumber
\end{align}

The changes in the translational and rotational velocity due to the disturbance viscosity are then
\begin{align}
    \boldsymbol{U}_1' & =  \frac{\eta_c}{12 \varepsilon \eta_{\infty}} \mathcal{Z}^{U'} \boldsymbol{U}_0 + \frac{a B_2}{30} (\mathcal{X}^{U',c} \mathsf{\t I} - \mathcal{Y}^{U',c} 3 \boldsymbol{pp}) \cdot {\boldsymbol{\nabla}} (\frac{\eta_0 }{ \eta_{\infty}}), \label{eqn:cv_disturbance_tran}\\
    \boldsymbol{\Omega}_1' & = \frac{1}{4} \mathcal{X}^{\Omega',c} \boldsymbol{U}_{N} \times {\boldsymbol{\nabla}} (\frac{\eta_0 }{ \eta_{\infty}}),\label{eqn:cv_disturbance_ang}
\end{align}
where
\begin{align*}
    \mathcal{Z}^{U'} & = \frac{6 \left(1 - e^2\right) \left(2 e - \mathcal{L}_e\right)^2}{e^2 \mathcal{L}_e [2e - \left(1 - e^2 \right) \mathcal{L}_e]}, \\
    \mathcal{X}^{U',c} & = 5 (-1 + e^2) \Big [ 8e^3(-9 -33e^2 + 32 e^4 ) - 4 e^2 \left( - 27 - 39e^2 + 62e^4\right) \mathcal{L}_e \\
    & \quad + 6e ( -9 + 7e^2 -3e^4 + 5e^6) \mathcal{L}_e^2 - 3 \left(-3 + 9e^2 - 13e^4 +7 e^6\right) \mathcal{L}_e^3 \Big ] \\
    & \quad \times \Big \{ 4 e^5 [6e + (-3+e^2)\mathcal{L}_e][2e+(-1+e^2)\mathcal{L}_e] \Big \}^{-1}, \\
    \mathcal{Y}^{U',c} & = 5  \Big [ 16e^4(27 - 27e^2 - 33 e^4 + 32 e^6 ) - 8 e^3 \left( 108 - 153e^2 -9 e^4 +62e^6\right) \mathcal{L}_e \\
    & \quad + 4e^2 ( 162 - 297e^2 + 147e^4 - 23e^6 + 15 e^8) \mathcal{L}_e^2 \\
    & \quad - 2e \left(108 - 243e^2 + 189e^4 - 77 e^6 + 23 e^8\right) \mathcal{L}_e^3  + 3(3-4e^2+e^4)^2 \mathcal{L}_e^4 \Big ] \\
    & \qquad \times \Big \{ 12 e^5(2e -\mathcal{L}_e) [6e + (-3+e^2)\mathcal{L}_e][2e+(-1+e^2)\mathcal{L}_e] \Big \}^{-1}, \\
    \mathcal{X}^{\Omega',c} & = \frac{ \left(1 - e^2\right) \left[ 4 e^2 (5 - 4e^2) + 20e(-1+e^2) \mathcal{L}_e + (5 - 6e + e^4) \mathcal{L}_e^2\right]}{\left(2 - e^2\right) [2e + (-1+e^2)\mathcal{L}_e]^2}. \\
\end{align*}

Adding \eqref{eqn:cv_disturbance_tran} and \eqref{eqn:cv_disturbance_ang} to \eqref{eqn:tran_vel} and \eqref{eqn:rot_vel} we obtain the combined effects of ambient and disturbance viscosities, on the particle's translational and rotational velocities
\begin{align}
    \boldsymbol{U}_1 & =  \frac{\eta_c}{12 \varepsilon \eta_{\infty}} \mathcal{Z}^{U} \boldsymbol{U}_0 - \frac{a B_2}{6} (\mathcal{X}^{U,c} \boldsymbol{I} - \mathcal{Y}^{U,c} 3 \boldsymbol{pp}) \cdot \boldsymbol{\nabla} \left(\frac{\eta_0 }{ \eta_{\infty}}\right),\label{eqn:constant_translational_velocity}\\
    \boldsymbol{\Omega}_1 & = - \frac{1}{4} \mathcal{X}^{\Omega,c} \boldsymbol{U}_{0} \times \boldsymbol{\nabla} \left(\frac{\eta_0 }{ \eta_{\infty}}\right), \label{eqn:constant_angular_velocity}
\end{align}
where
\begin{align}
\allowdisplaybreaks
    \mathcal{Z}^{U} & = \frac{6 \left(1 - e^2\right) \left(2 e - \mathcal{L}_e\right)^2}{e^2 \mathcal{L}_e [2e - \left(1 - e^2 \right) \mathcal{L}_e]}, \\
    \mathcal{X}^{U,c} & =  \Big [ -4e^2(45 - 75e^2 + 32 e^4 ) + 12 e (15 - 28e^2 + 13e^4) \mathcal{L}_e - 9 (1 - e^2)^2 (5 + e^2) \mathcal{L}_e^2 + 6e  (1 - e^2)^2 \mathcal{L}_e^3 \Big ] \nonumber \\
    & \qquad \times \Big \{ 2 e^2 [6e + (-3+e^2)\mathcal{L}_e][2e + (-1+e^2)\mathcal{L}_e] \Big \}^{-1}, \\
    \mathcal{Y}^{U,c} & =  \Big [ -4e^4(-39 + 32e^2 ) -  \left( 168 e^3 - 156 e^5\right) \mathcal{L}_e  - e^2 ( -63 + 54e^2 + 5e^4 ) \mathcal{L}_e^2 - e (15 - 16e^2 + e^4)  \mathcal{L}_e^3 + 3(1 - e^2)^2 \mathcal{L}_e^4 \Big ] \nonumber \\
    & \qquad \times \Big \{ 3 e [6e + (-3+e^2)\mathcal{L}_e](2e - \mathcal{L}_e)[2e+(-1+e^2)\mathcal{L}_e] \Big \}^{-1}, \\
    \mathcal{X}^{\Omega,c} & = \frac{ \left(1 - e^2\right) \left[ 4 e^2 (- 7 + 4e^2) - 4e(-7 + 5e^2) \mathcal{L}_e + (-7 + 6e^2 + e^4) \mathcal{L}_e^2\right]}{\left(2 - e^2\right)  [2e - (1-e^2)\mathcal{L}_e]^2}.
\end{align}
Compared to the no-flux condition, the disturbance viscosity here introduces a more complex influence on the swimming dynamics of a spheroidal particle. However, the particles will still generally display viscophobic dynamics.

\bibliography{spheroids}

\end{document}